\documentclass[review,12pt,nopreprintline]{elsarticle}

\usepackage{amssymb}
\usepackage{graphicx}
\usepackage{stix}

\usepackage{siunitx}
\usepackage{multirow}
\usepackage{tabularx} 
\usepackage[table]{xcolor}
\usepackage{pdflscape}
\definecolor{LightCyan}{rgb}{0.88,1,1}
\definecolor{OrangeViolin}{rgb}{0.88186275, 0.50539216, 0.17303922}
\definecolor{BlueViolin}{rgb}{0.19460784, 0.45343137, 0.63284314}
\definecolor{NiceRed}{rgb}{0.9019607843137255, 0.09803921568627451, 0.29411764705882354}
\definecolor{NiceGreen}{rgb}{0.23529411764705882, 0.7058823529411765, 0.29411764705882354}
\definecolor{NiceYellow}{rgb}
{1.0, 0.8823529411764706, 0.09803921568627451}
\usepackage{subcaption}

\journal{Artificial Intelligence in Medicine}

\begin{document}

\begin{frontmatter}



\title{Coronary artery segmentation in non-contrast calcium scoring CT images using deep learning}


\author[GLI]{Mariusz Bujny}
\ead{mbujny@graylight-imaging.com}

\author[GLI,US]{Katarzyna Jesionek}
\ead{kjesionek@graylight-imaging.com}

\author[GLI,SUT]{Jakub Nalepa}
\ead{jnalepa@graylight-imaging.com}

\author[GLI,SCCS]{Karol Miszalski-Jamka}
\ead{kmiszalski@graylight-imaging.com}

\author[GLI]{Katarzyna Widawka-Żak}
\ead{kwidawka@graylight-imaging.com}

\author[GLI]{Sabina Wolny}
\ead{swolny@graylight-imaging.com}

\author[GLI,US]{Marcin Kostur\corref{cor}}
\ead{mkostur@graylight-imaging.com}

\cortext[cor]{Corresponding author}

\affiliation[GLI]{organization={Graylight Imaging},
            addressline={ul. Bojkowska 37a}, 
            city={Gliwice},
            postcode={44-100}, 
            state={Silesian Voivodeship},
            country={Poland}}

\affiliation[US]{organization={University of Silesia},
            addressline={ul. Bankowa 12}, 
            city={Katowice},
            postcode={40-007}, 
            state={Silesian Voivodeship},
            country={Poland}}

\affiliation[SUT]{organization={Silesian University of Technology},
            addressline={ul. Akademicka 16}, 
            city={Gliwice},
            postcode={44-100}, 
            state={Silesian Voivodeship},
            country={Poland}}

\affiliation[SCCS]{organization={Silesian Center for Heart Diseases},
            addressline={ul. Marii Skłodowskiej-Curie 9}, 
            city={Zabrze},
            postcode={41-800}, 
            state={Silesian Voivodeship},
            country={Poland}}

\begin{abstract}
Precise localization of coronary arteries in Computed Tomography (CT) scans is critical from the perspective of medical assessment of coronary artery disease. Although various methods exist that offer high-quality segmentation of coronary arteries in cardiac contrast-enhanced CT scans, the potential of less invasive, non-contrast CT in this area is still not fully exploited. Since such fine anatomical structures are hardly visible in this type of medical images, the existing methods are characterized by high recall and low precision, and are used mainly for filtering of atherosclerotic plaques in the context of calcium scoring. In this paper, we address this research gap and introduce a deep learning algorithm for segmenting coronary arteries in multi-vendor ECG-gated non-contrast cardiac CT images which benefits from a novel framework for semi-automatic generation of Ground Truth (GT) via image registration. We hypothesize that the proposed GT generation process is much more efficient in this case than manual segmentation, since it allows for a fast generation of large volumes of diverse data, which leads to well-generalizing models. To investigate and thoroughly evaluate the segmentation quality based on such an approach, we propose a novel method for manual mesh-to-image registration, which is used to create our test-GT. The experimental study shows that the trained model has significantly higher accuracy than the GT used for training, and leads to the Dice and clDice metrics close to the interrater variability.
\end{abstract}


\end{frontmatter}


\section{Introduction}
\label{sec:introduction}
Segmentation of coronary arteries in Computed Tomography (CT) scans plays a critical role in the assessment and treatment of coronary artery disease, which is one of the leading causes of death worldwide~\cite{go_executive_2013, sun_computational_2014}. The availability of coronary artery models reconstructed based on CT with contrast enhancement, frequently referred to as Coronary Computed Tomography Angiography (CCTA), allows for utilization of computational techniques in a variety of problems, including blood flow simulation and biomechanical analysis~\cite{sun_computational_2014, malawski_deep_2022}, virtual stenting and stent testing~\cite{beier_hemodynamics_2016, sun_personalized_2019}, and it enables building virtual reality as well as physical models for educational purposes~\cite{silva_emerging_2018, yoo_hands-surgical_2017}. 
The effort required for traditional, manual segmentation of coronary arteries makes it impractical in clinical applications due to the unacceptable amount of analysis time it demands. 
With the advent of deep learning  in medical imaging in recent years~\cite{isensee_nnu-net_2021, wasserthal_totalsegmentator_2022}, fast, robust, and fully automatic segmentation of coronary arteries became possible~\cite{gharleghi_automated_2022}, which opens a wide range of new research pathways for applying various computational techniques in clinical practice.

\begin{figure}[ht!]
\centering
\includegraphics [width=1.0\textwidth]{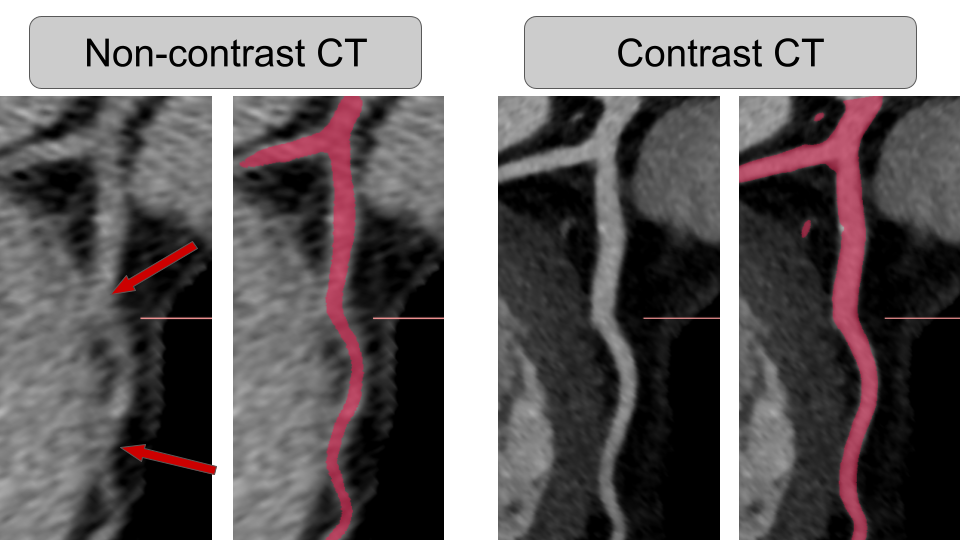}
\caption{
The challenge of segmenting the coronary vessel from non-contrast CT is demonstrated by an example of the Left Anterior Descending (LAD) artery visualized in Stretched Curvilinear Reformulation (CPR)---the same vessel is shown in non-contrast and contrast CT. In each case, we show the CT image alone (left) and an overlaid ML segmentation mask (right). The red arrows point to the areas where Hounsfield values of the vessel and surrounding tissue are practically the same, which makes applying classic segmentation techniques, like region growth methods, impossible. Note that the Hounsfield range for non-contrast and contrast CT is different, and is set to $(-120,200)$ and $(-120,800)$, respectively. }
\label{fig:nosee}
\end{figure}

The state-of-the-art techniques for segmenting coronary arteries in CCTA commonly exploit 2D and 3D U-Net architectures~\cite{gharleghi_automated_2022}, which have been successfully used to solve a variety of problems in medical image segmentation~\cite{siddique_u-net_2021}. The best-performing methods to date~\cite{isensee_nnu-net_2021} yield mean Dice scores up to 0.88 over the unseen test scans~\cite{gharleghi_automated_2022}. 
Nevertheless, a challenge persists to accurately segment distal parts of coronary arteries and maintain topological correctness of the resulting coronary vessel tree models, which is neither sufficiently reflected in the most common loss functions used in the network training nor in the standard segmentation quality metrics. In order to thoroughly quantify the capabilities of emerging algorithms, an array of metrics have been suggested. Apart from standard segmentation metrics such as Dice score, Recall, and Precision, specialized metrics for vessel segmentation have been proposed, including centerline Dice (clDice)~\cite{Shit_2021_CVPR} as well as the corresponding clPrecision and clRecall.



One of the most important challenges in the field is accurate segmentation of coronary arteries in non-contrast CT scans, used so far mainly for calcium scoring~\cite{agatston_quantification_1990}, \textcolor{black}{but, if done more precisely, having a great potential in terms of improving downstream tasks such as assignment of calcifications to proper branches of the coronary artery tree. Moreover, by solving such a problem, one could significantly improve explainability of the automatic calcium scoring methods via utilization of visualization techniques taking advantage of additional anatomical information. Potentially, Machine Learning (ML) models based on non-contrast CT could offer diagnostic value for patients with health contradictions to contrast agent administration, as well.} \textcolor{black}{Nevertheless, such a segmentation task is particularly} demanding due to limited information about \textcolor{black}{coronary arteries, being very} fine-grained structures\textcolor{black}{,} in an image without contrast enhancement~\cite{isgum_automatic_2012, wolterink_automatic_2014, yang_automatic_2016, lessmann_automatic_2018}. Even for experienced medical imaging specialists, it is challenging to model the main branches of the vessel tree based on such images using the commonly available segmentation tools. In Fig.~\ref{fig:nosee}, we demonstrate this difficulty in an example of a well-visible Left Coronary Artery (LCA) vessel fragment in non-contrast CT containing no imaging artifacts. Even in this case, the coronary vessel merges with the surrounding tissue in certain regions (indicated by the arrows in Fig.~\ref{fig:nosee}), making local signal thresholding insufficient for accurate segmentation. Thus, the segmentation algorithm must incorporate data extrapolation and leverage the atlas' knowledge of the vascular tree anatomy to ensure precise vessel geometry segmentation. We address this research gap of automated coronary artery segmentation in non-contrast CT \textcolor{black}{by proposing a novel framework for automatic generation of GT data and training ML models based on weak supervision by a human expert who performs only fast, binary decisions to assess the quality of training data, which is the main contribution of this work. We demonstrate the value of this process by introducing a new, manual mask-to-image registration method and applying it to a set of cases, including open-source data, to generate unique GT samples for testing, which we make publicly available.} \textcolor{black}{In Section~\ref{sec:contribution},} the most important contributions of our work are \textcolor{black}{are discussed in more detail}, whereas the current state of the art in the field is \textcolor{black}{summarized} in Section~\ref{sec:sota}.

\subsection{Coronary artery segmentation in non-contrast CT}
\label{sec:sota}



The approaches toward automatic segmentation of coronary arteries in non-contrast CT can be split into two groups, (\textit{i})~using image registration, including selected atlas-based approaches, and (\textit{ii}) those which exploit ML, where deep neural networks play an essential role. The first group of methods takes advantage of the fact that the coronary arteries are relatively well visible in contrast-enhanced CT. Since in the standard protocol for atherosclerosis assessment both contrast and non-contrast series are acquired~\cite{shahzad_patient-specific_2010}, it is possible to use the methods for coronary artery segmentation in contrast CT, and align the resulting model with the non-contrast image using registration techniques~\cite{klein_elastix_2010, shamonin_fast_2013, joshi_unbiased_2004}. When only a non-contrast CT scan is available, a generic vessel tree model can be used to segment the coronary arteries in the non-contrast image directly. In one of the first approaches of this type, Shahzad et al.~\cite{shahzad_patient-specific_2010, shahzad_vessel_2013} used 85 contrast CT scans with manual annotations of the main coronary arteries' centerlines to construct a set of 10 atlases via nonrigid transformation of images. As a result, each of the atlases contained voxel density information indicating the location of coronary vessels. Finally, each of the atlas images was aligned with the non-contrast scan based on automatic registration using Elastix~\cite{klein_elastix_2010}, and the resulting density information was averaged to give a rough estimation of the location of the main coronary arteries. A similar, multi-atlas approach was used by Wolterink et al.~\cite{wolterink_automatic_2014}. Here, a small set of contrast CT images with manual annotations of LAD, Left Circumflex (LCX), and Right Coronary Artery (RCA), were aligned with the test image by means of affine transformation, which was followed by elastic registration using Elastix as well~\cite{klein_elastix_2010}. The transformations were used to propagate the coronary vessel tree models from the atlases, and consensus centerlines were estimated using a Selective and Iterative Method for Performance Level Estimation (SIMPLE)~\cite{langerak_label_2010}. In an alternative approach, Kelm et al.~\cite{kelm_automatic_2014} used a robust centerline tracing algorithm~\cite{zheng_robust_2013} to extract patient-specific coronary centerline vessel trees from contrast CT scans. At the same time, in both contrast and non-contrast series, pericardium and the aortic root were segmented using the marginal space learning algorithm~\cite{zheng_four-chamber_2008}. The masks were used to fit the coronary centerline tree to the non-contrast image via registration using the thin-plate-spline model~\cite{zheng_pericardium_2014}. Then, Kondo et al.~\cite{kondo_semi-automatic_2015} performed a patient-specific registration of the main coronary arteries, segmented in contrast CT, to the non-contrast CT. However, the authors used a semi-automatic process utilizing region growing techniques to obtain the segmentations in the contrast CT scans and relied on a different registration framework~\cite{johnson_brainsfit_2007}. Finally, Yang et al.~\cite{yang_automatic_2016} also used a patient-specific image registration consisting of consecutive affine and nonrigid transformations. The coronary arteries were automatically segmented based on the original contrast CT scan using Frangi's vesselness filter~\cite{frangi_multiscale_1998}. To increase the accuracy of the coronary artery segmentation, after the standard image-to-image registration, nonrigid registration was performed exclusively within the estimated heart region. All in all, despite the high practical value of the registration-based coronary vessel segmentation techniques, which helped to significantly improve the calcium scoring process, the segmentation accuracy is relatively low~\cite{lee_fully_2021}, especially in the context of the recent developments in deep learning for medical image analysis~\cite{suganyadevi_review_2022, chen_recent_2022, isensee_nnu-net_2021, malawski_deep_2022, kotowski_detecting_2023}. Moreover, compared to deep learning solutions, atlas-based methods are considerably slower and the computation time increases substantially for a high number of atlases~\cite{lee_fully_2021}. Additionally, methods using patient-specific registration require availability of both contrast and non-contrast CT series. The above mentioned drawbacks of the registration-based methods motivate the development of deep learning approaches for segmenting coronary arteries in non-contrast CT.

The second group of methods consists of approaches which benefit from classic ML relying on handcrafted feature extractors as well as deep neural networks that exploit automated representation learning~\cite{isensee_nnu-net_2021, wasserthal_totalsegmentator_2022}. Brunner et al.~\cite{brunner_toward_2010} proposed to use affine transformations of the scan coordinate system to obtain a common representation of the heart region across different patients. The reference frame was transformed to match predefined anatomical landmarks, being the aortic root and the apex of the heart. Then, using transformed coordinates as the input features, a support vector machine was trained on manually labeled trajectory points of the main coronary artery zones, i.e., LAD with Left Main (LM) coronary artery, LCX, and RCA, where each of the zones was divided into three equidistant subsections. The model was not used directly to segment the coronary arteries, but was utilized for labeling of coronary vessel calcifications. The method, however, is not fully automatic, since it requires localization of the anatomical landmarks by a human expert. Gonzalez et al.~\cite{gonzalez_automated_2016} proposed an automated Agatston score computation methodology based on a ML method for identifying the beginnings of the coronary arteries. The great vessels of the heart, e.g., aorta, pulmonary artery, and coronaries, were detected utilizing 2D object detectors, being adaptive boosting classifiers using Haar wavelets as the input features. The segmented organs were used to enhance the detection of coronary artery calcifications based on standard thresholding~\cite{agatston_quantification_1990} by defining a region of interest corresponding to the location of the heart. However, it is difficult to judge the accuracy of detecting the coronary arteries. Also, the method uses exclusively 2D images, which is a severe limitation in medical image analysis. Finally, Lee et al.~\cite{lee_fully_2021} proposed a promising approach using both image registration and 3D deep learning techniques for image segmentation. For each of 100 patients, contrast CT images with manual segmentations of coronary arteries and other surrounding structures including, i.a., aorta, left and right ventricle, and mitral annulus, were aligned with the corresponding non-contrast scans using image registration based on Elastix~\cite{klein_elastix_2010, shamonin_fast_2013}. The deformed masks of organs in the reference frame of non-contrast CT scans, together with the images, were used as GT for training 3D-patch U-Nets~\cite{simonyan_very_2015}. For detection of coronary arteries, a patch of $64\times64\times64$ voxels of size of 1 [mm] in each direction was used, and the masks were dilated two times with $2\times2\times1$ morphological structuring elements to account for registration errors. The information about the location of the main branches of the coronary vessel tree, i.e., RCA, LAD, LCX, LM, was included in GT as well. The resulting deep network was able to learn an anatomical atlas model of coronary arteries, which was used for detecting coronary artery calcifications. Although the accuracy metrics for coronary artery segmentation were not presented, the results for calcium scoring were better than in the other works~\cite{kurkure_supervised_2010, shahzad_vessel_2013, wolterink_automatic_2015}, which suggests good segmentation accuracy. Furthermore, the method allows to reduce the computational time compared to the multi-atlas approaches~\cite{shahzad_patient-specific_2010, shahzad_vessel_2013, wolterink_automatic_2014}. The segmentation model, however, was tailored for detecting coronary calcium and used dilated, low-resolution masks covering larger regions around the coronary arteries, and it is not suitable for tasks requiring higher level of precision, such as lumen segmentation. Moreover, it seems that the outputs of the U-Net model were post-processed to include only the two largest components, which might result in an increase of false negative classifications, especially for the cases including severe calcifications and artifacts.

\subsection{Motivation and contribution of this work}
\label{sec:contribution}



The available approaches based on ECG-gated non-contrast CT mainly focus on calcium scoring rather than coronary artery segmentation itself, and therefore, usually provide only low-resolution segmentation masks. Here, precise coronary tree models obtained from non-contrast CT images could be directly utilized to develop improved algorithms that automatically segment calcification within coronary arteries, thereby enhancing their effectiveness not only in calculating the total coronary artery calcium score but also in improving the assignment of calcification to the appropriate segment of the coronary artery tree. Currently available software on commercial diagnostic consoles still does not provide automated solutions to those tasks. Additionally, visualization of coronary arteries in non-contrast CT images can also be employed for the detection of potential anomalies in the origin and course of the coronary arteries. Furthermore, promising efforts are currently underway to develop advanced algorithms utilizing radiomics for the assessment of coronary artery plaques and pericoronary adipose tissue, which are based on manual segmentations of coronary arteries on ECG-gated non-contrast CT scans~\cite{Jiang2022}. Therefore, full automation of the segmentation process of non-contrast CT images could enhance the performance of these algorithms and pave the way for broader utilization of novel markers for cardiovascular risk assessment.

\begin{figure*}[ht!]
\centering
\includegraphics [width=\textwidth]{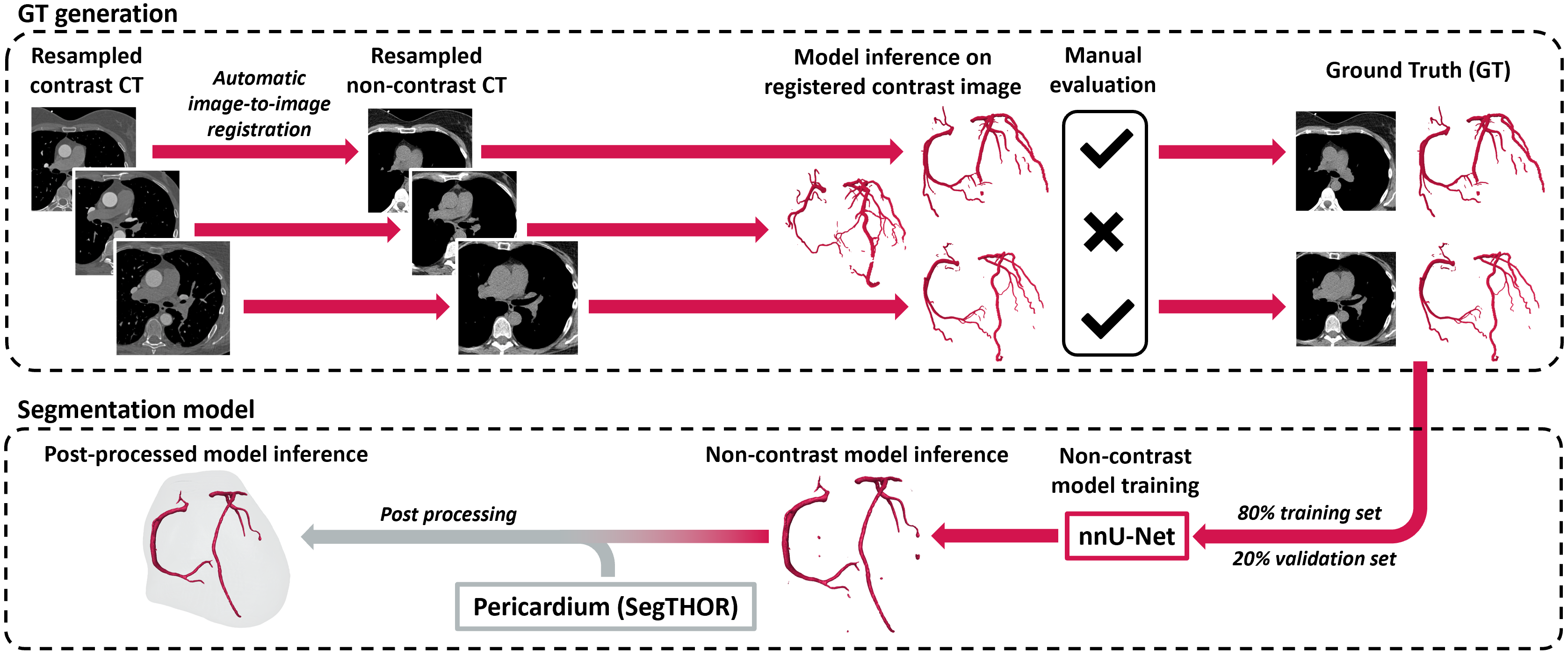}
\caption{Framework for semi-automatic GT generation and training of non-contrast coronary artery segmentation models.}
\label{fig:method_overview}
\end{figure*}

The objectives of this work are multi-fold. The primary objective is thus to propose a scalable end-to-end framework (Fig.~\ref{fig:method_overview}) for automated, high-quality segmentation of coronary arteries in multi-vendor ECG-gated non-contrast cardiac CT scans, and to thoroughly evaluate its accuracy. To ensure full reproducibility of experiments, we build our implementation on publicly available state-of-the-art image registration~\cite{yushkevich_fast_2016} and deep learning~\cite{isensee_nnu-net_2021} techniques. Moreover, we propose a novel methodology for the generation of GT for testing the resulting ML model based on manual registration of high-resolution coronary artery masks segmented in contrast CT to the corresponding non-contrast CT images. This allows for computation of both standard and specialized segmentation quality metrics, which we confront with the metrics from an interrater evaluation. The results of our experimental study performed over a stratified test set consisting of 17 multi-vendor ECG-gated non-contrast cardiac CT scans with the manually registered coronary models confirm high potential of the proposed deep learning approach, which effectively learns an atlas able to interpolate the segmentation in the areas without a clear signal in the image. Based on qualitative analysis, we conclude that the obtained segmentations are significantly more accurate than the ones presented in the literature, opening up possibilities for their use in other application areas.

The remainder of this paper is structured as follows. Section \ref{sec:segmentation_framework} presents our coronary artery segmentation framework. In Section \ref{sec:manual_registration}, the manual mesh-to-image registration method for generation of \textit{test-GT} is proposed. Section \ref{sec:materials} summarizes the dataset used in this study. In Section \ref{sec:results}, a description of the training process and the qualitative and quantitative evaluation of the segmentation quality is presented. Section \ref{sec:conclusion} concludes the work and gives indications for the future research.

\section{Registration-based deep learning framework for coronary artery segmentation}
\label{sec:segmentation_framework}

In this section, a semi-automatic framework for GT generation and deep learning model training is described. Our approach is inspired by Lee et al.~\cite{lee_fully_2021}, however, we exploit the state-of-the-art open-source algorithms~\cite{yushkevich_fast_2016, isensee_nnu-net_2021} with default parameters to allow for an easy reimplementation of the proposed method and replication of results. In contrast to Lee et al.~\cite{lee_fully_2021}, our research prioritizes high Precision in segmentation models over Recall, as our aim is to reconstruct the anatomy of coronary vessels, rather than to simply facilitate calcium scoring. Therefore, we suggest to use upsampled images and apply additional verification steps by medical imaging experts to generate GT for training of the neural network model. Fig.~\ref{fig:method_overview} presents an overview of the proposed framework, whose main components are discussed in detail in the sections below. 


\subsection{Generation of GT for training deep learning models}
\label{sec:registration}


Due to the challenging character of manual segmentation of coronary arteries in non-contrast CT, which prohibits the creation of large datasets required by deep learning techniques, we propose to semi-automatically generate the training GT based on a three-step approach. For each patient, we perform an image-to-image registration by moving the contrast scan to the reference frame of the non-contrast CT. This first step is realized within the Greedy framework\footnote{https://github.com/pyushkevich/greedy, accessed on 25\textsuperscript{th} May 2023.} \cite{yushkevich_fast_2016, joshi_unbiased_2004} with default parameters. For the sake of generality, we rely in the further processing exclusively on the registered images. Hence, the registration process can be treated as a black box, and other algorithms~\cite{keszei_survey_2017}, including commercial solutions, can be easily used here as well.
Then, we propose to use a model of coronary arteries trained on contrast CT to segment coronary vessels in the reference frame of the non-contrast image based on the registered contrast scans. We exploit our 3D U-Net model~\cite{malawski_deep_2022} built upon the nnU-Net framework~\cite{isensee_nnu-net_2021}. However, thanks to the generality of the proposed approach, any other model~\cite{wasserthal_totalsegmentator_2022} can be easily applied to automatically generate GT candidates.


Finally, the candidate GT samples consisting of non-contrast images and the corresponding coronary artery segmentation models are reviewed by a human expert. We suggest selecting a subset of samples that satisfy specific quality criteria, evaluated based on visual inspection of a registered coronary vessel mask overlaid on a non-contrast image. Here, as the evaluation criterion, we took the segmentation visual accuracy of the proximal parts of coronary arteries. Those parts are usually clearly visible in non-contrast CT images, as they do not ``touch'' the myocardium (see the upper part of Fig.~\ref{fig:nosee}).



\subsection{Deep learning segmentation model}
\label{sec:deep_learning_model}
We utilize the state-of-the-art nnU-Net framework~\cite{isensee_nnu-net_2021} for training of a coronary artery segmentation model based on the semi-automatically generated GT. The nnU-Net automatically configures its hyperparameters and network structure during the training process, and delivers an optimized 3D U-Net model. Both non-contrast CT scans and segmentation masks are in the upsampled isotropic spacing of 0.35\,mm. Due to the demanding hardware requirements of the high-resolution coronary vessel segmentation, the final segmentation is generated based on network inferences in multiple patches of the $160 \times 160 \times 96$ size (in voxels). The output models of coronary arteries are post-processed to eliminate the artifacts generated by the neural network. This is accomplished by filtering out all the groups of connected voxels of volume less than 50\,mm\textsuperscript{3} and removing all the voxel groups which are positioned entirely outside of pericardium, which we segment in non-contrast CT using an openly available SegTHOR model~\cite{lambert_segthor_2020}.

\section{Manual mesh-to-image registration method for generation of test-GT}
\label{sec:manual_registration}
Since the deep learning models presented in this paper are trained based exclusively on the semi-automatically generated GT, the evaluation of the segmentation quality plays a critical role, and a separate GT for testing is needed to thoroughly understand the generalization capabilities of the elaborated models. However, using standard tools based on region-growing techniques to generate \textit{test-GT} is not feasible due to low image intensity gradients in the CT scans without contrast enhancement (Fig.~\ref{fig:nosee}). Hence, we propose a novel method for generating \textit{test-GT} via manual registration of geometric models of coronary arteries to non-contrast CT scans. Here, we describe a process based on the tools available in Blender\footnote{https://www.blender.org, accessed on 25\textsuperscript{th} May 2023.}, an open-source 3D computer graphics software widely used in various domains. However, the presented concepts can be easily transferred to other graphics software packages as well.

We introduce a manual mesh-to-image registration method which operates as follows. First, given a 3D surface mesh of the coronary vessel tree being registered, i.e., LCA or RCA, a centerline representation is automatically extracted using skeletonization techniques~\cite{saha_survey_2016} or similar approaches~\cite{izzo_vascular_2018}. Here, we used the Vascular Modeling Toolkit (VMTK)\footnote{http://www.vmtk.org, accessed 25\textsuperscript{th} May 2023.}, which is an open-source software package specialized for the geometric processing of blood vessels. The resulting representation can contain artifacts, e.g., extra centerlines, hence, the models should be manually post-processed in the 3D computer graphics software to remove the additional objects.

\begin{figure}[ht!]
\centering
\includegraphics [width=0.5\textwidth]{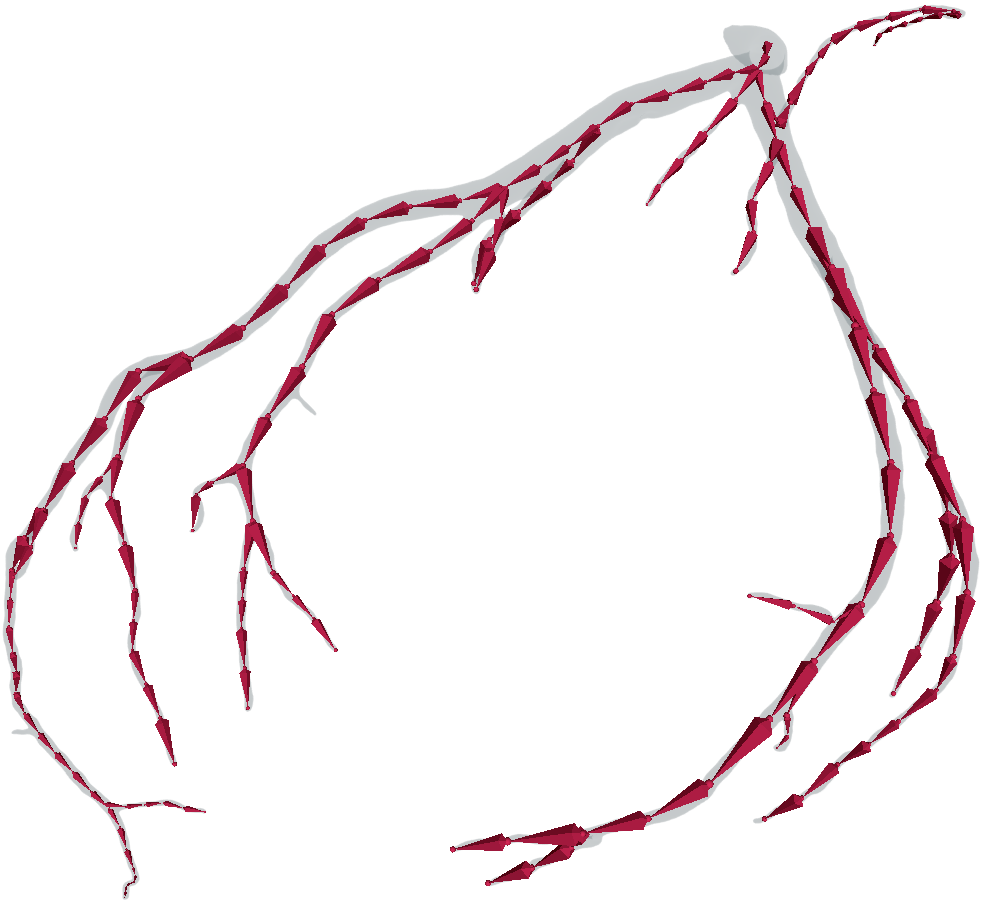}
\caption{High-resolution LCA tree model with its skeletal representation in Blender. The tips and roots of individual \textit{Bones} coincide with the vertices of the centerlines.}
\label{fig:lca_bones}
\end{figure}

\begin{figure}[ht!]
\centering
\begin{subfigure}{0.21\textwidth}
\includegraphics[width=\textwidth]{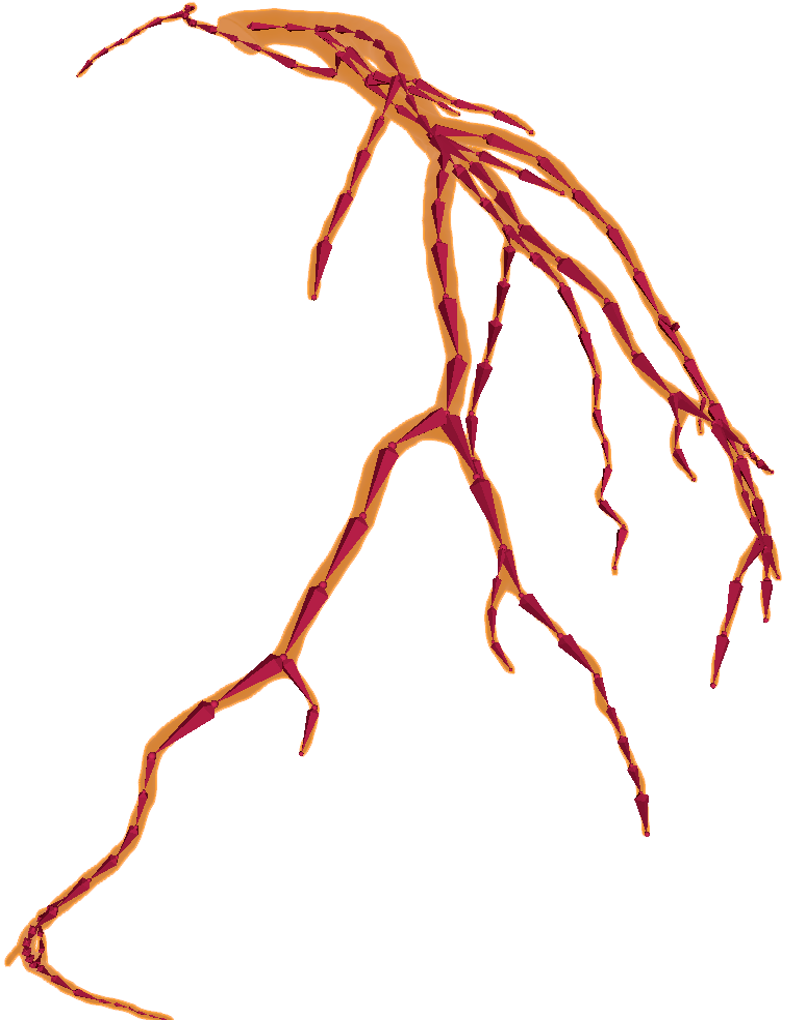}
\caption{Initial mesh.}
\end{subfigure}
\begin{subfigure}{0.31\textwidth}
\includegraphics[width=\textwidth]{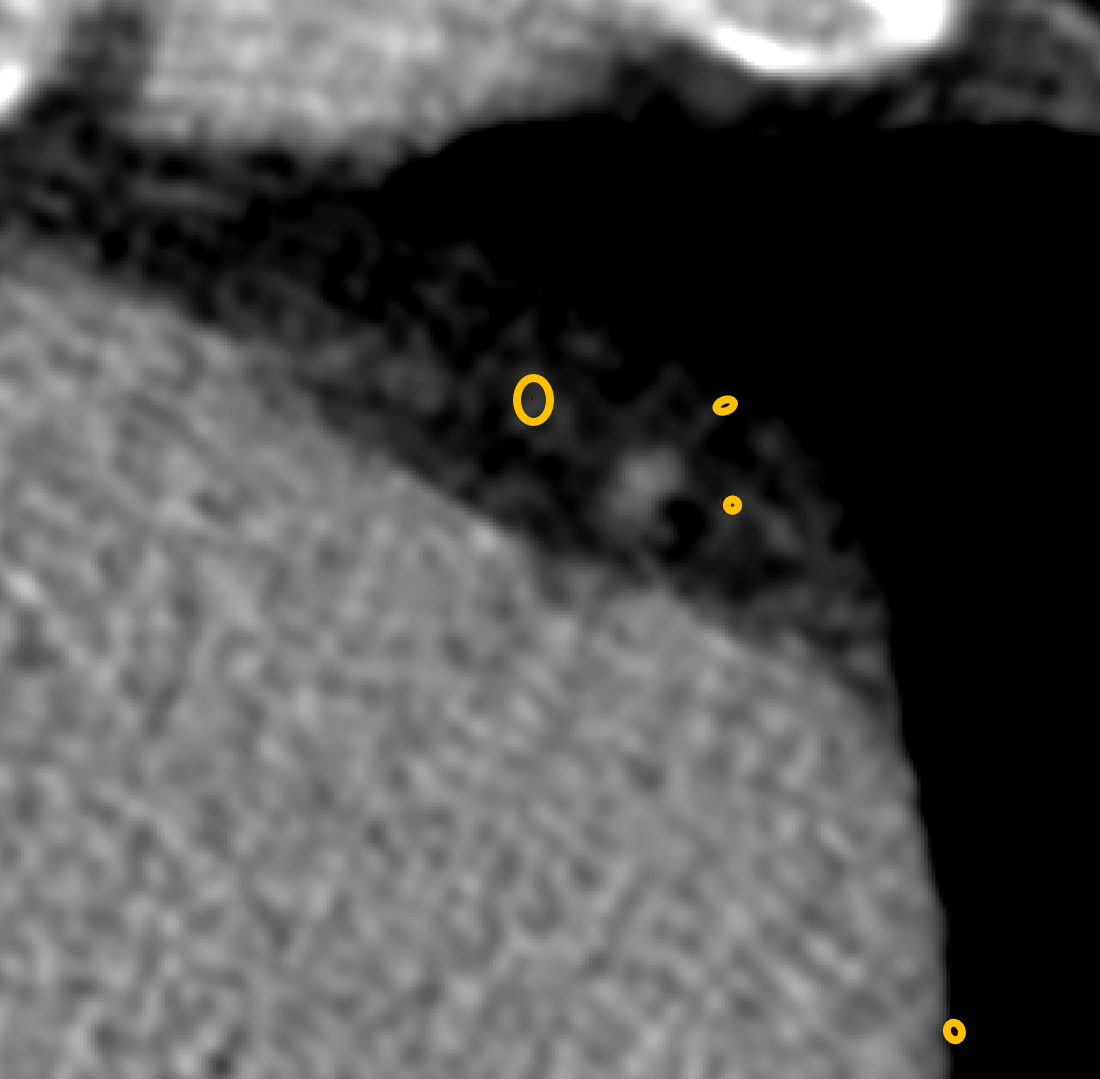}
\caption{Axial view (initial LCA contours).}
\end{subfigure}
\newline
\centering
\begin{subfigure}{0.21\textwidth}
\includegraphics[width=\textwidth]{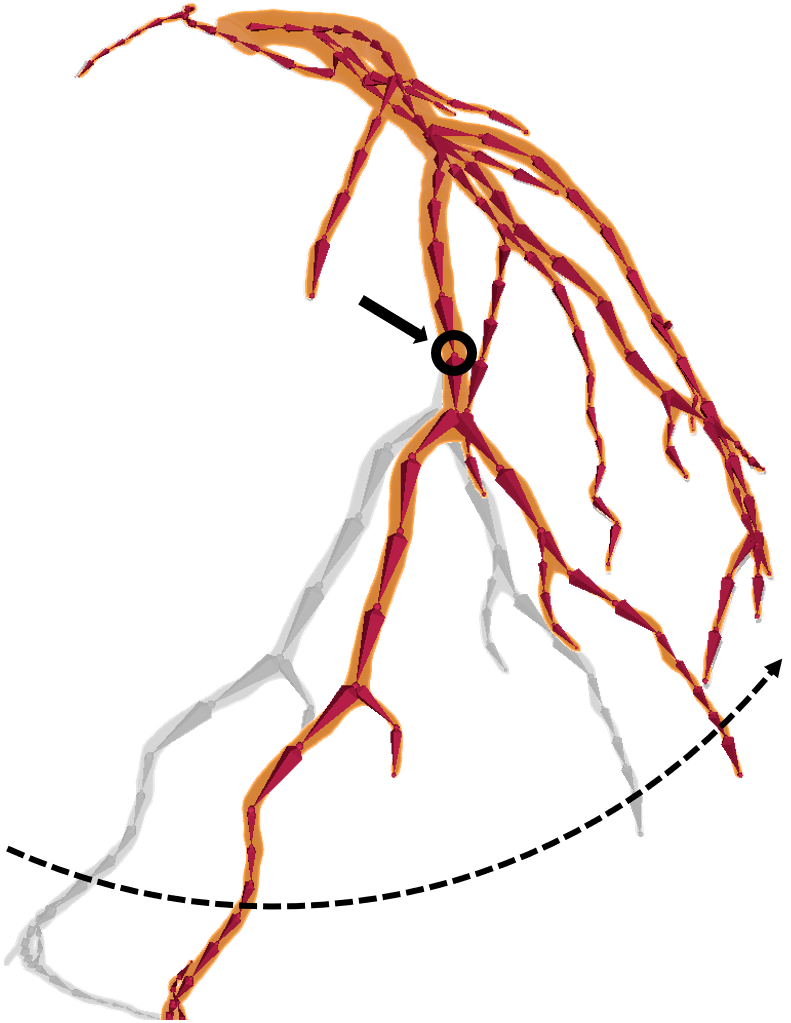}
\caption{Step 1.}
\end{subfigure}
\begin{subfigure}{0.31\textwidth}
\includegraphics[width=\textwidth]{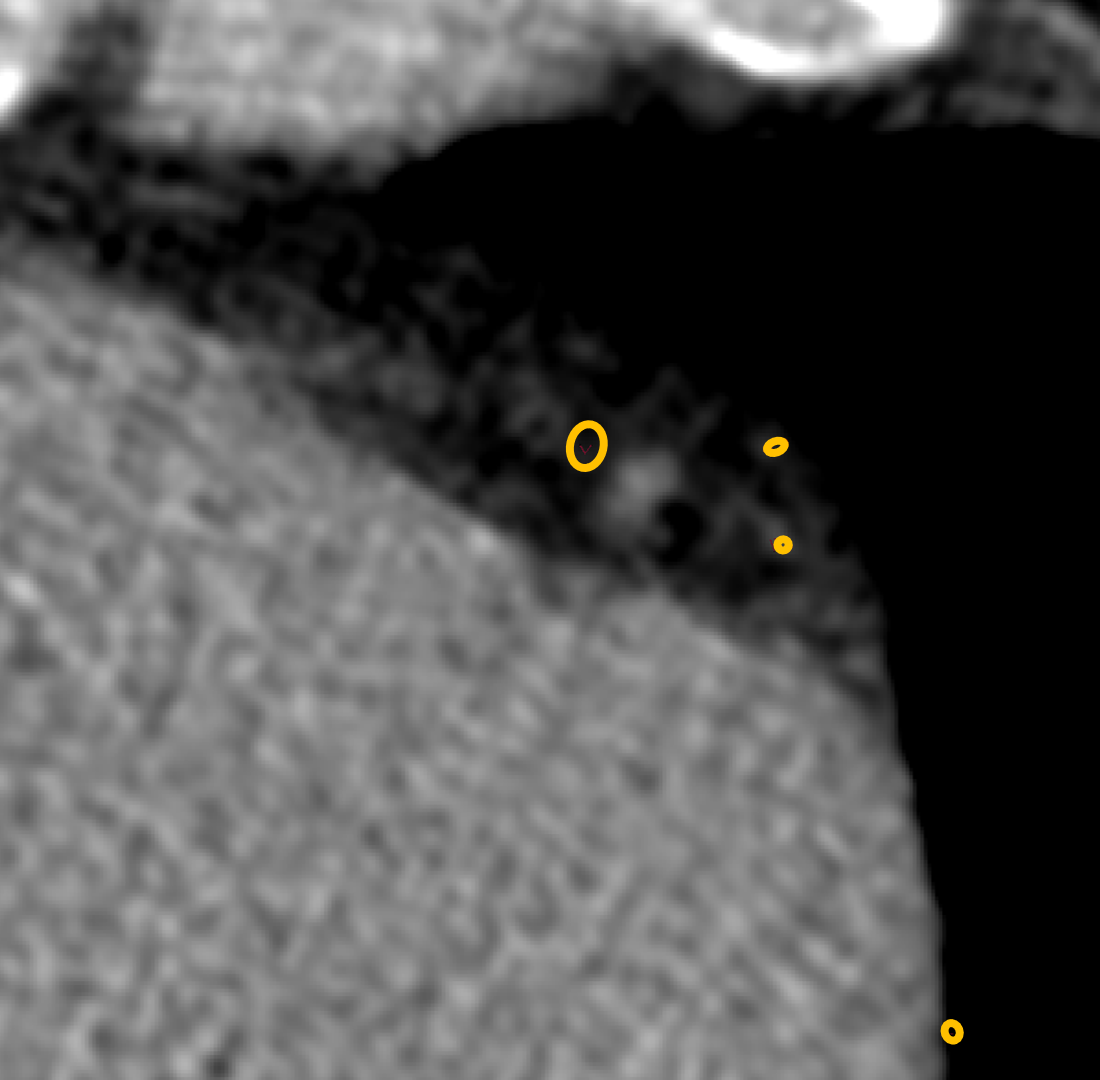}
\caption{Axial view, LCA contours (Step 1).}
\end{subfigure}
\newline
\centering
\begin{subfigure}{0.21\textwidth}
\includegraphics[width=\textwidth]{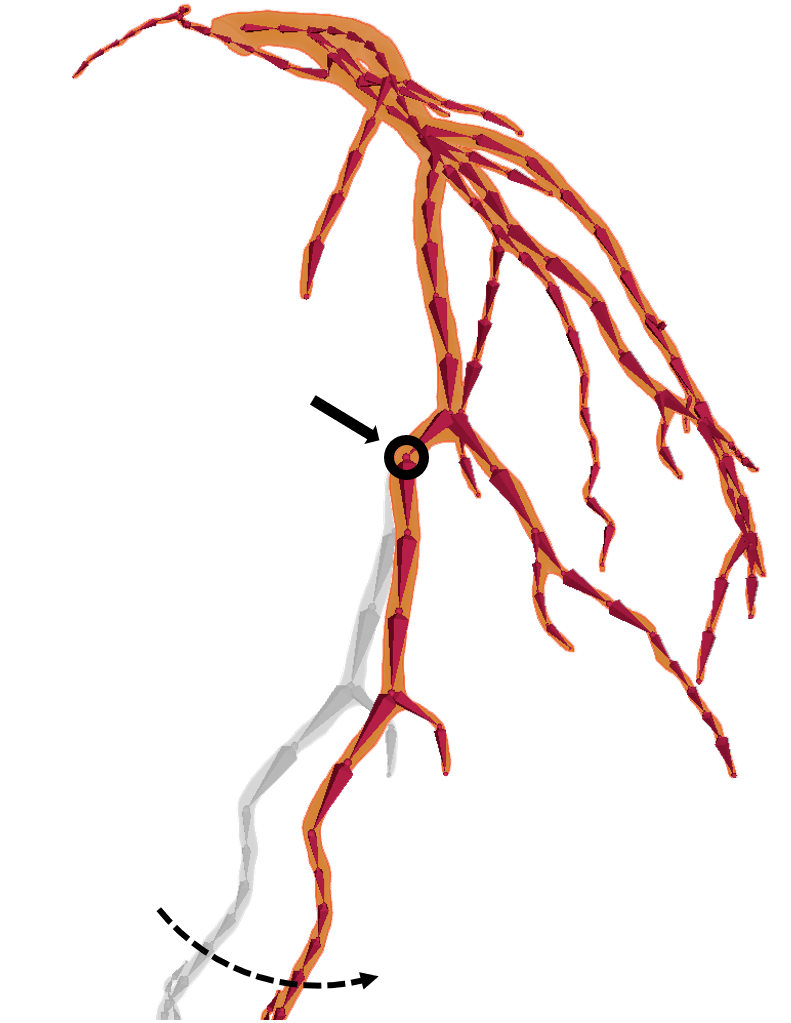}
\caption{Step 2.}
\end{subfigure}
\begin{subfigure}{0.31\textwidth}
\includegraphics[width=\textwidth]{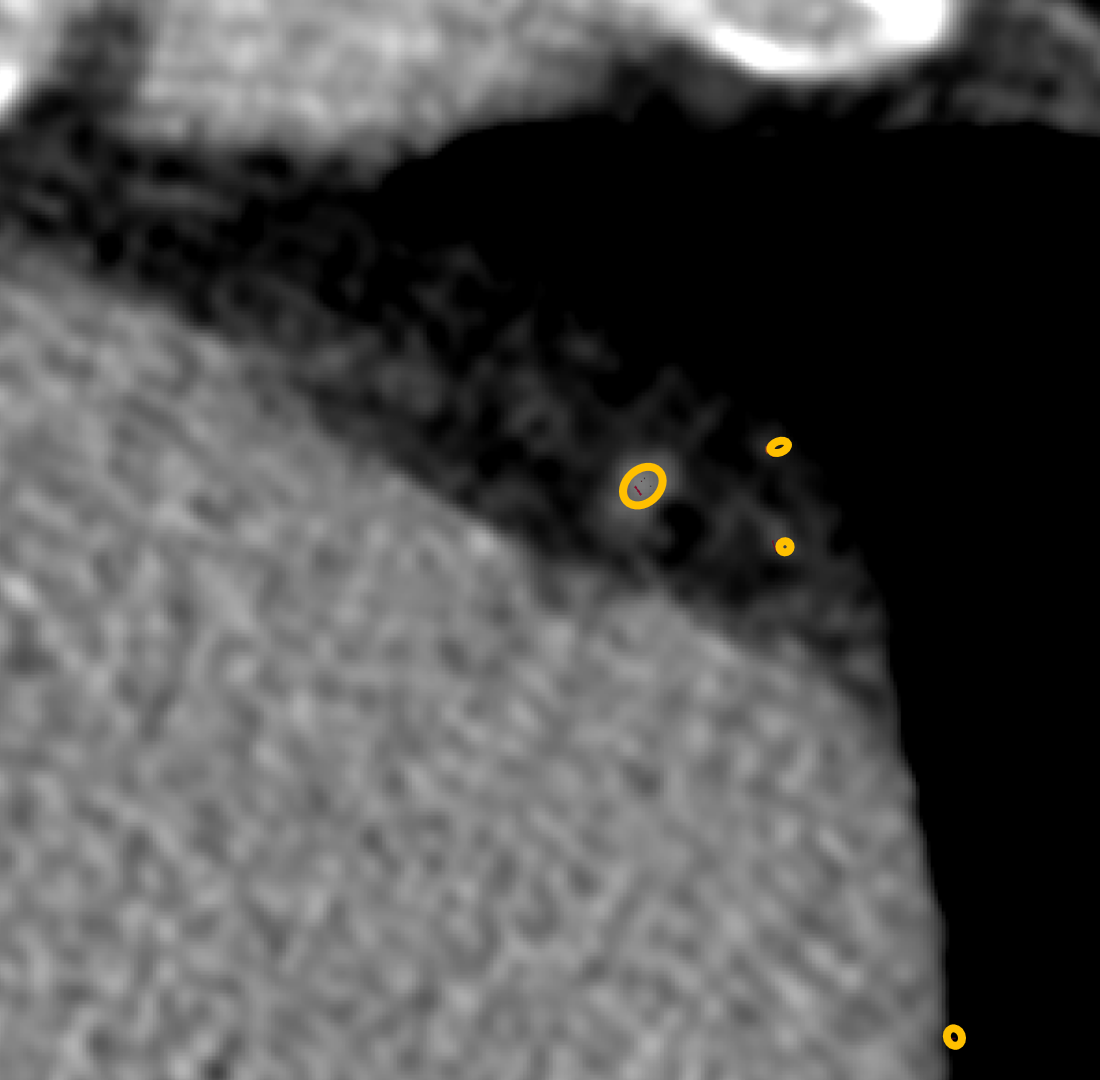}
\caption{Axial view, LCA contours (Step 2).}
\end{subfigure}
\newline
\caption{The alignment of a 3D coronary vessel tree model (panels (a), (c), and (e)) with the non-contrast CT scan (panels (b), (d), and (f)) based on rotations of subsequent \textit{Bones}. The black solid arrows and circles in the 3D views (a), (c), and (e) indicate the center of rotation at a given alignment step. The rotation directions are shown using dashed curves. The corresponding contours of the LCA tree at a fixed slice of the CT scan are depicted in the axial views (b), (d), and (f).}
\label{fig:bones_rotation}
\end{figure}

Subsequently, based on the extracted centerlines, we create a skeletal representation consisting of the Blender's \textit{Bone}\footnote{https://docs.blender.org/manual/en/2.79/rigging/armatures/bones, accessed on 25\textsuperscript{th} May 2023.} objects, which will serve as a geometric parameterization of the original, high-resolution 3D mesh model. To achieve this, \textit{Bones} are created along consecutive sections of each centerline, by aligning their roots and tips with the curve vertices using the \textit{Snap} functionality, as depicted in Fig.~\ref{fig:lca_bones}. We recommend using \textit{Bones} of approximately 3\,mm length or shorter, if a refinement of the representation is needed due to high curvature of a centerline. In Blender, we propose to use the \textit{Extrude} tool to quickly generate a single \textit{Chain of Bones} for the entire coronary vessel tree. At $n$-furcation points, the \textit{Extrusion} operation should be used more than once to create separate branches belonging to the same \textit{Chain of Bones}. As a result, the downstream \textit{Bones} will move in a coordinated manner with the \textit{Bones} located in the upstream branches when the latter ones are displaced. It is of note that for the sections where multiple centerlines overlap, the \textit{Bone} objects should be created only once. Finally, a relationship between the generated skeletal representation and the high-resolution vessel tree model has to be established. In Blender, this can be done using \textit{Armature Deform Parenting}, e.g., \textit{With Automatic Weights}. As a result, the influence of each \textit{Bone} on the mesh to deform is calculated automatically according to the distance from the \textit{Bone} to the vertices of the mesh.

Based on the newly created skeletal parameterization of the coronary vessel tree, the actual manual registration process can be carried out. This is done by iteratively rotating tips of the \textit{Bones} to match the contours of the tree model with the vessels visible in the non-contrast CT scan based on an axial view\footnote{To display the axial slices of the CT scan together with the contours of the coronary vessel tree, we used the Cardiac Add-on for Blender, a specialized module for 3D geometric modeling of coronary arteries developed at Graylight Imaging (https://graylight-imaging.com/3d-modelling). However, the entire procedure of manual mesh-to-image registration can be realized using exclusively standard Blender's tools, e.g., multiple \textit{Plane} objects with textures corresponding to the axial slices of a CT scan, but it would be an even more time-consuming process.}, as shown in Fig.~\ref{fig:bones_rotation}. The process starts by applying a rigid transformation of the initial vessel tree model to align its root with the ostium, which is usually clearly visible in a non-contrast CT scan. Subsequently, starting from the root of the tree, individual \textit{Bones} are rotated as illustrated in Fig.~\ref{fig:bones_rotation}. The deformations of the vessel tree are very intuitive since the child branches of the \textit{Bone} being modified are rotated by the same angle, which is only a rigid transformation of the downstream part of the model. Once a part of the branch is no longer visible in the non-contrast scan, it is cut at that point, and all of its child branches are removed from the model.

The manual mesh-to-image process is finalized by applying local corrections of the shape of the tree, e.g., with use of \textit{Grab} and \textit{Thumb} brushes in Blender. Finally, the resulting, post-processed coronary vessel tree models are stored and used as \textit{test-GT} in the evaluation presented in Section \ref{sec:quantitative_evaluation}.

\section{Materials}
\label{sec:materials}

To train a model capable of segmenting coronary arteries in ECG-gated non-contrast CT scans according to the procedure presented in Section \ref{sec:deep_learning_model}, a set of contrast-enhanced and the corresponding non-contrast CT scans is required to generate GT semi-automatically, as described in Section \ref{sec:registration}. Hence, in the section below, both types of CT scans, as well as the geometric models of coronary vessels resulting from the semi-automatic segmentation process are described. Moreover, to evaluate the capabilities of the trained model, a dataset consisting of patient-specific segmentations of coronary arteries based on contrast CT and the corresponding non-contrast CT scans is required. Following the mesh-to-image registration method presented in Sect.~\ref{sec:manual_registration}, we use this data to create a \textit{test-GT} dataset, which is described in the following section.




\subsection{Selection of CT scans}

The selection of samples comprising the datasets used in this study for training of the models was based on the qualitative assessment of segmentation quality (as described in Section \ref{sec:registration}) by a medical imaging specialist with 4 years of clinical experience. Additionally, scans with anatomical abnormalities or high noise levels were excluded according to the specialist's visual inspection. The final result of the selection procedure was a dataset of 98 multi-vendor ECG-gated cardiac non-contrast CT scans with corresponding masks of coronary vessels. We refer to the data corresponding to each patient as a \emph{case}. The manually prepared \textit{test-GT} (Section \ref{sec:manual_registration}) was based on a randomly chosen set of 17 multi-vendor scans from a separate pool of cardiac CT scans, each containing both contrast and non-contrast series (Table~\ref{table:datasets}). A subset of 9 cases, 3 in the training and 6 in \textit{test-GT} sets, were derived from the OrCaScore dataset~\cite{wolterink_evaluation_2016}, being a public set of calcified plaque scores from the MICCAI 2014 Challenge on Automatic Coronary Calcium Scoring.

\begin{table}[ht!]
\centering
\scriptsize
\caption{Characteristics of the CT scans used in this study.}
\label{table:datasets} 
\begin{tabularx}{0.96\linewidth}{|X|l|l|}
 \hline 
 \rowcolor{black!7!}

\textbf{Characteristic}&  \textbf{Training dataset}&  \textbf{\textit{test-GT} dataset}   \\  \hline
 \rowcolor{LightCyan}
\hspace{0mm}Population characteristics&  &  \\ \hline
\hspace{2mm} Age, years& $59\pm11$ & $64\pm9$   \\  \hline
\hspace{2mm} Male &  35\% & 18\%  \\ \hline
 \rowcolor{LightCyan}
\hspace{0mm}CT scans& &  \\ \hline
\hspace{2mm}\textbf{Scanner model}& &  \\ \hline

\hspace{3mm} Siemens SOMATOM\newline \hspace*{3mm} Definition AS+ & 3 &  2 \\ \hline
\hspace{3mm} Siemens SOMATOM\newline \hspace*{3mm} Definition Flash & 92 & 10  \\ \hline
\hspace{3mm} Siemens SOMATOM \newline \hspace*{3mm} Force & 0 &  2 \\ \hline
\hspace{3mm} GE Lightspeed VCT & 1 &  3 \\ \hline
\hspace{3mm} Philips Brilliance iCT & 2 &  1 \\ \hline
\hspace{3mm} Toshiba Aquilion ONE & 0 &  2 \\ \hline
\rowcolor{red!20!green!15!}
\hspace{2mm}Non-contrast CT scan& &  \\ \hline
\hspace{2mm}\textbf{Tube voltage, [kVp]}& &  \\ \hline
\hspace{3mm}$120$& 98 & 17 \\ \hline
\hspace{2mm}\textbf{In-plane voxel size, [mm]} &
$0.41 \pm 0.04$& $0.44\pm0.08$ \\ \hline
\rowcolor{red!20!green!15!}
\hspace{2mm}CCTA scan& &  \\ \hline
\hspace{3mm}\textbf{Tube voltage, [kVp]}& &  \\ \hline
\hspace{4mm}$80$& 0 & 3 \\ \hline
\hspace{4mm}$100$& 91 & 13 \\ \hline
\hspace{4mm}$120$& 5 & 1 \\ \hline
\hspace{4mm}$140$& 2 & 0 \\ \hline
\hspace{2mm}\textbf{In-plane voxel size, [mm]} &
$0.41 \pm 0.04$& $0.40\pm0.04$ \\ \hline

\end{tabularx}
\end{table}



\subsection{GT segmentations}
In this study, we exploit two separate GT sets: a semi-automatically generated GT (Section \ref{sec:registration}) for training deep learning models (98 segmentations of coronary arteries), alongside the \textit{test-GT} set for evaluating their generalization capabilities. To generate the \textit{test-GT}, a group of experienced medical imaging specialists manually segmented coronary arteries in the contrast CT scans. Then, a set of 17 coronary tree segmentations was created by a medical imaging expert (4 years of experience), by aligning the contrast-based segmentations with the non-contrast CT scans according to the procedure described in Section \ref{sec:manual_registration}. As a result, a \textit{test-GT} set consisting of pairs of non-contrast CT scans and the aligned coronary vessel segmentations for 17 patients was created. To allow for an easy inspection of the resulting segmentations, we publish~\cite{bujny_dataset_2023} a subset of 6 \textit{test-GT} samples generated based on the scans belonging to the OrCaScore dataset~\cite{wolterink_evaluation_2016}. Additionally, to assess the quality of the manual alignment procedure, a second rater (medical expert, 2 years of clinical experience) performed additional alignments on a randomly selected set of 10 vessel trees. These segmentations were used in the quantitative evaluation of our model (Section \ref{sec:quantitative_evaluation}).






\begin{landscape} 
\begin{table*}[htb]
\scriptsize
\caption{Coronary artery segmentation metrics obtained by the deep model over the \textit{validation} and \textit{test-GT} sets, including the impact of post-processing (\textit{vol50}: filtered out all groups of connected voxels of volume less than 50 mm\textsuperscript{3}, \textit{pericardium}: removed all voxel groups which lay entirely outside of pericardium)---the results obtained using the final deep learning processing chain are boldfaced. The metrics for interrater test 
(\textit{test-GT} vs. second rater) are shown as well.}
\label{tab:metrices_test}
\centering
\begin{tabularx}{1.4\textwidth}{|c|c|c|c|c|c|c|c|c|c|c|}
 \hline
  \rowcolor{LightCyan}
  \multicolumn{2}{|X|}{}
  &\multicolumn{3}{|c|}{\textbf{Dice}}&
  \multicolumn{3}{|c|}{\textbf{Recall}}&
  \multicolumn{3}{|c|}{\textbf{Precision}}
  \\
 \hline
 \textbf{Compared segmentations}&
 \textbf{Post-processing}&
 \textbf{mean}&
 \textbf{median} &
 \textbf{std. dev.}&
 \textbf{mean}&\textbf{median} &\textbf{std. dev.}&\textbf{mean}&\textbf{median} &\textbf{std. dev.}\\
  \hline
 \rowcolor{blue!5!black!7!green!20}
  Interrater test&
None &0.67&0.67&0.09&0.67&0.65&0.11&0.68&0.68&0.07\\
  \hline
  \rowcolor{red!5!black!7!orange!20}
  ML model vs. \textit{validation set} &None&0.57&0.55&0.10 &0.55&0.53&0.11&0.61&0.60&0.10\\
  \hline
\multirow{4}{*}{ML model vs. \textit{test-GT}}&\textbf{vol50+pericardium}&
\textbf{0.65}
&\textbf{0.65}
&\textbf{0.08}
&\textbf{0.61}
&\textbf{0.59}
&\textbf{0.10}
&\textbf{0.73}
&\textbf{0.76}
&\textbf{0.11}
 \\ \cline{2-11}
  &vol50&0.64
&0.64
&0.08
&0.61
&0.59
&0.10
&0.71
&0.74
&0.13
\\
  \cline{2-11}
&pericardium&
0.65
&0.64
&0.07
&0.63
&0.62
&0.10
&0.68
&0.69
&0.10
\\
  \cline{2-11}
 
 &None &0.62 
 &0.63&0.09
 &0.63
 &0.62
 &0.09
 &0.63
 &0.65
 &0.13
 \\
  \hline
  \hline
  \rowcolor{LightCyan}
  \multicolumn{2}{|c|}{}&\multicolumn{3}{|c|}{\textbf{clDice}}&\multicolumn{3}{|c|}{\textbf{clRecall}}&\multicolumn{3}{|c|}{\textbf{clPrecision}}\\
 \hline
 \textbf{Compared segmentations}&
 \textbf{Post-processing}
 &\textbf{mean}&\textbf{median} &\textbf{std. dev.}
 &\textbf{mean}&\textbf{median} &\textbf{std. dev.}
 &\textbf{mean}&\textbf{median} &\textbf{std. dev.}\\
 \hline
 \rowcolor{blue!5!black!7!green!20}
Interrater test& 
None&0.64&0.63&0.06&0.64&0.57&0.1&0.65&0.66&0.06\\
\hline
  \rowcolor{red!5!black!7!orange!20}
ML model vs. \textit{validation set} & None &0.43&0.41&0.11& 0.34 & 0.31 & 0.10& 0.59 & 0.58 & 0.15\\
  \hline
\multirow{4}{*}{ML model vs. \textit{test-GT}}&\textbf{vol50+pericardium}
&\textbf{0.65}
&\textbf{0.63}
&\textbf{0.08}
&\textbf{0.52}
&\textbf{0.49}
&\textbf{0.11}
&\textbf{0.91}
&\textbf{0.95}
&\textbf{0.08}
\\ \cline{2-11}
& vol50
&0.64
&0.63
&0.06
&0.52
&0.48
&0.11
&0.87
&0.93
&0.13
\\ \cline{2-11}
&pericardium
&0.65
&0.65
&0.07
&0.56
&0.52
&0.09
&0.79
&0.81
&0.07
\\ \cline{2-11}

&None 
&0.62
&0.62
&0.10
&0.56
&0.53
&0.10
&0.71
&0.74
&0.14
\\
\hline

\end{tabularx}
\end{table*}
\end{landscape}

\section{Results}
\label{sec:results}
In this section, we report and discuss the results of our experimental study. Section~\ref{sec:training} describes the training process of the network for segmentation of coronary arteries in non-contrast CT scans. In Section~\ref{sec:quantitative_evaluation}, we present a critical, qualitative as well as quantitative analysis of the deep model.

\subsection{Training a deep learning model}\label{sec:training}

The dataset of 98 cases (Section \ref{sec:materials}) was used for training a deep neural network within the nnU-Net framework, where 20 CT scans were used exclusively for validation of the model during the training procedure. The training process was carried out until reaching nnU-Net's default limit of 1000 epochs. The quality metrics for the inferences of the resulting model on the \textit{validation set} (20 cases) are presented in Table \ref{tab:metrices_test}.





\begin{figure}[ht!]
\centering
\includegraphics[width=1.0\textwidth]{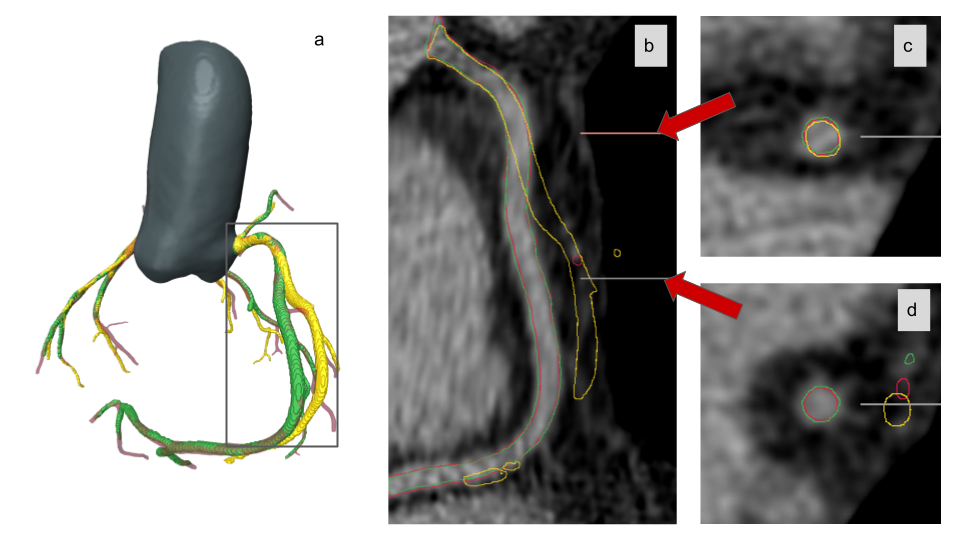}
\caption{Example of a mismatch between the GT created based on registration ({\color{NiceYellow}$ \mdblksquare $}), the inference of the neural network ({\color{NiceGreen}$ \mdblksquare $}), and the manually aligned contrast GT ({\color{NiceRed}$ \mdblksquare $}) in the distal part of RCA (a), juxtaposed with the corresponding non-contrast CT scan.
Panel (b) is a CPR projection along the vessel centerline indicated in (a) by a solid black rectangle. Panels (c) and (d) show cuts in the vessel's short axis at two distinct points. Based on the visual comparison, we can note that the inference  of the deep model and manually aligned vessel are consistent with the image, unlike the GT used in the training process.}
\label{fig:validation_gt_mismatch}
\end{figure}

With a mean \textit{validation set}'s Dice at the level of $0.57 \pm 0.10$, the segmentation accuracy is relatively low compared to the best methods for segmentation of coronary arteries based on contrast CT, reaching Dice scores up to 0.88~\cite{gharleghi_automated_2022}\footnote{Although we are aware that confronting different algorithms over different datasets (and training-test splits) may easily become misleading, we can indeed observe that the Dice scores are significantly larger for contrast CT scans.}. However, there are two important aspects which should be noted here. First, for coronary arteries, being very thin structures, the Dice metric decreases substantially even in case of small errors on the voxel level, which also results in low interrater Dice for this type of segmentation problems. The information contained in the non-contrast CT scan is far more scarce than in contrast CT, which naturally leads to higher segmentation errors as well. Secondly, and more importantly, the metrics presented here are calculated on a \textit{validation set} consisting of semi-automatically generated GT. The sole criterion used in the selection of CT scans accepted in our study for the training set and the \textit{validation set} was the alignment of the proximal regions of the coronary arteries. Therefore, the lower Dice scores observed in this case may be attributed to the lower quality of the GT data in the distal regions. This issue is illustrated in Fig.~\ref{fig:validation_gt_mismatch}, which shows that segmentation of the distal part of RCA based on the trained model is more consistent with the CT scan than the GT used for training. Indeed, in some cases, the inference of
the model is better than the underlying GT used for both training and validation, which is counter-intuitive, but can be justified by the strong regularization imposed on the U-Net model. The network learns robust filters to detect vessels from those parts of the dataset where registration is correct and is not degraded by misinformation caused by massive but relatively rare misalignments. Fig.~\ref{fig:validation_gt_mismatch} is a prominent example that the model is more accurate than the GT used for its training.


Even though a deep learning model trained on imperfect data may produce visually appealing segmentation results, it is important to accurately evaluate its performance using reliable GT data. To address this issue, we manually created the \textit{test-GT} dataset, as detailed in Section \ref{sec:manual_registration}. The evaluation results based on \textit{test-GT} are presented in the following section.




\subsection{Quantitative evaluation based on test-GT}
\label{sec:quantitative_evaluation}

\begin{figure}[ht!]
\centering
\includegraphics[width=0.75\textwidth]{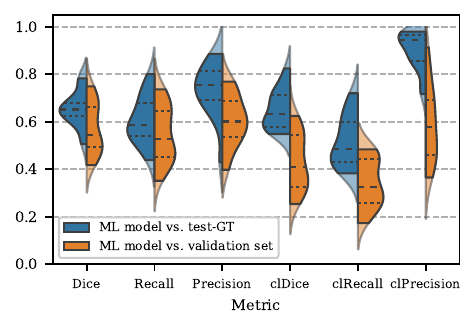}
\caption{Distribution of the metrics obtained by the deep learning (ML) model over the \textit{test-GT} set ({\color{BlueViolin}$ \mdblksquare $}) and the ML model tested on the \textit{validation set} 
({\color{OrangeViolin}$ \mdblksquare $}).
The non-transparent areas 
indicate the sample range. Additionally, the $25\%$ and 75\% percentiles ({\textbf{\texttt{$\cdot${}$\cdot$}}}) and median ({\texttt{-{}-}}) are marked in the plot.}
\label{fig:V_test_val}
\end{figure}

We utilize a manually aligned \textit{test-GT} which contains information regarding the actual anatomy of the patient's coronary vessels to assess the model quality. Classic segmentation measures, including Dice, Precision, and Recall, along with their centerline counterparts, clDice, clPrecision, and clRecall, are computed. These metrics are presented in Fig.~\ref{fig:V_test_val} and gathered in Table~\ref{tab:metrices_test}. A subset of 6 inferences based on the scans from the OrCaScore dataset~\cite{wolterink_evaluation_2016} is available online~\cite{bujny_dataset_2023}\footnote{Note that we publish the inferences for this subset of \textit{test-GT} only due to the fact that the corresponding CT scans are openly available. As a result, they might not be fully representative of our test cases.}.

The primary outcome of this work is an ML model for the segmentation of coronary arteries in non-contrast CT scans, yielding Dice coefficient of $0.65\pm0.08$ on the manually registered \textit{test-GT}. Fig.~\ref{fig:V_test_val} summarizes these results and compares the distribution of the metrics obtained on the \textit{validation set} as well as \textit{test-GT}. The evaluation reveals that the performance of the model computed on the \textit{test-GT} is consistently better than when computed over the \textit{validation set}, which is a consequence of the significant but rare misalignment during image-to-image registration. This observation is confirmed by Mann-Whitney tests verifying if the differences in distributions of the metrics are statistically significant. For Dice, Precision, clDice, clRecall and clPrecision they are indeed statistically significant with the $p$-values of 0.0305, 0.0008, $<0.0001$, $<0.0001$, and $<0.0001$, respectively. For Recall, the $p$-value amounted to 0.1582, showing that this metric remained consistent for both datasets, with significantly better Precision for \textit{test-GT}.


While the overlap metrics (Dice, Precision, and Recall) are close to each other on both datasets, the metrics describing the coronary tree, i.e., clDice, clPrecision, and clRecall, are significantly higher when computed over the \textit{test-GT} set. In case of the segmentations based on our ML model, the Dice coefficient is a geometric average of a low Recall and a high Precision. To correctly understand why Recall is low, let us note that it is intrinsically limited by the low visibility of coronary vessels in non-contrast CT data. At the same time, we compare the model-based segmentations with the tree geometries derived from high-quality contrast CT scans. Conversely, high Precision indicates that if the model detects a vessel, it will likely be the actual vessel of a patient. The degree of overlap between coronary trees is quantified by the clPrecision metric, which achieves values exceeding 0.9.

\begin{figure}[ht!]
\centering
\includegraphics [width=1.0\textwidth]{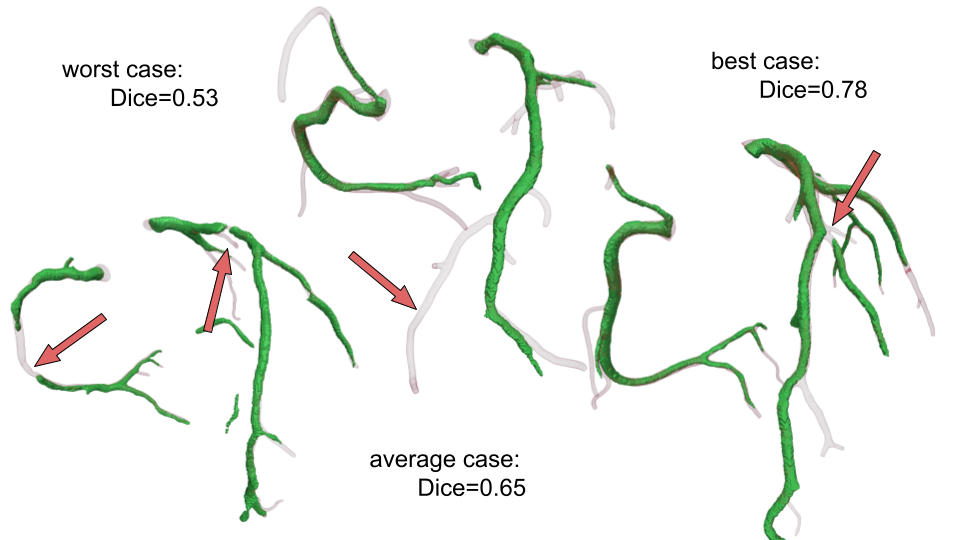}
\caption{Inferences obtained using the ML model for three selected \textit{test-GT} (green) samples and the corresponding, manually registered coronary vessel trees (light gray). The worst (Dice below average), average, and the best (Dice above average) cases are presented. The ML model correctly predicts vessel geometry in all cases, and the results coincide with the manually aligned GT. However, in all cases, many small branches are missing. Red arrows indicate model's imperfections such as small discontinuities and missing branches.
}
\label{fig:examples}
\end{figure}

Examining the model's inferences on a per-test-case basis can provide valuable insights. Here, we present only three extreme cases: the worst, medium, and the best inferences of the deep learning model on \textit{test-GT}, as quantified by Dice. In Fig.~\ref{fig:examples} we can appreciate that the basic coronary tree structure is recovered in all the cases. The worst case, with Dice equal 0.53, has a significant discontinuity in the RCA tree, caused by a motion artifact in the CT acquisition. In all other cases, small and easily repairable discontinuities are present. One should also note that the source of the low Recall and clRecall values lies primarily in missing, hardly visible vessel endings.

The nnU-Net uses a standard sliding window for CT image processing, which results in some false positives as patches often do not have information regarding the global heart position. Thus, it is reasonable to use post-processing techniques to eliminate small disconnected vessel fragments (the \textit{vol50} post-processing routine) and structures outside the pericardium (\textit{pericardium} post-processing), which, in this study, was segmented using another nnU-Net model, SegTHOR~\cite{lambert_segthor_2020}. Our study reveals that the post-processing (\textit{vol50+pericardium}) does significantly affect Dice, Recall, Precision, clRecall and clPrecision metrics (the $p$-values obtained using two-side Wilcoxon tests amounted to 0.0002 for Dice, and $<0.0001$ for all other metrics)---this systematic improvement can be also appreciated in Table \ref{tab:metrices_test}. It is of note that the combined post-processing (\textit{vol50+pericardium}) offered statistically significant improvements (at $p<0.05$) when compared with other post-processing variants involving either \textit{vol50} or \textit{pericardium} post-processing, as well as the lack of post-processing, in the majority ($11/18$) of pairwise comparisons and metrics. Therefore, all of the results presented in the analysis above included the (\textit{vol50+pericardium}) post-processing step. 

\begin{figure}[ht!]
\centering
\includegraphics[width=0.75\textwidth]{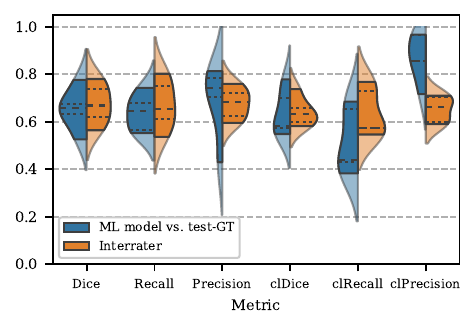}
\caption{Distribution of the interrater metrics ({\color{OrangeViolin}$ \mdblksquare $}) and those obtained by the ML model tested on the subset of \textit{test-GT} dataset containing the same five cases ({\color{BlueViolin}$ \mdblksquare $}).
The non-transparent areas 
indicate the sample range. Additionally, the $25\%$ and $75\%$ percentiles ({\textbf{\texttt{$\cdot${}$\cdot$}}}), and median ({\texttt{-{}-}}) are marked.}
\label{fig:V_test_inter}
\end{figure}

Finally, we approached the problem of assessing the quality of the manual alignment procedure. 
The results including a comparison with a second rater are shown in Fig.~\ref{fig:V_test_inter}, demonstrating that the procedure includes a certain degree of randomness. It can be noted that the second rater scores Dice and clDice slightly but systematically higher than the ML model compared to the first rater. However, in these cases, Precision and Recall are similar to each other.
Based on these findings, we conclude that the model's performance is on par with the interrater agreement, indicating that it likely extracts all the information that can be extracted from non-contrast CT scans, considering the image resolution and signal level.


\section{Conclusion and Outlook}
\label{sec:conclusion}
Segmentation of coronary arteries based on non-contrast cardiac computed tomography scans is a challenging medical image analysis task. However, using CT without contrast enhancement can be highly beneficial for patient diagnosis, as it is less invasive than contrast-based examinations. This paper proposed a method for achieving high-precision segmentation of coronary arteries based on multi-vendor ECG-gated non-contrast cardiac CT images, using a framework for semi-automatic Ground Truth (GT) generation (via image registration) and a deep neural network model. The effectiveness of the proposed solution was confirmed through a quantitative analysis over the test set generated using a novel manual mesh-to-image registration method, which was ultimately exploited to evaluate the quality of our approach.

The major finding of this work is that our deep neural network model is able to effectively generalize over the unseen multi-vendor ECG-gated non-contrast cardiac CT scans, and it allows for achieving high-quality results despite the relatively low-quality GT used for training. It is of clinical importance, as the segmentation of coronary arteries in low-dose non-contrast calcium scoring CT images can be utilized to generate more advanced algorithms for automatic coronary artery calcium scoring, leading to improved accuracy of coronary artery calcification detection and their assignment to the appropriate coronary tree segment. More precise segmentation of coronary arteries on non-contrast cardiac CT scans can also aid in identifying coronary anomalies and potentially enhance the performance of radiomics algorithms in evaluating coronary artery plaques and pericoronary adipose tissue \cite{Jiang2022}. 
Moreover, in future studies utilizing high-resolution ECG-gated non-contrast cardiac CT images, it could be tested whether deep learning coronary artery segmentation allows for the extraction of additional clinically relevant information from these images. Such an approach could be implemented initially in selected patient subgroups, such as those with contraindications to intravenous iodine-based contrast agent administration.

To further increase the segmentation accuracy, it is necessary to improve the quality of GT. Manual generation of GT seems unreasonable due to the time-consuming and user-dependent nature of this process. Improving the registration method is the key to increasing the quality of the semi-automatically generated GT and, consequently, creating better segmentation models. Another promising approach might be to eliminate the post-processing step and incorporate it into the machine learning algorithm via utilization of specialized techniques, e.g., graph neural networks. Modifying the model's architecture may also lead to an increase in the connectivity of the obtained segmentations. A greater challenge, however, is how to improve the model's ability to segment the vessel endings, which are practically invisible in the non-contrast CT scans. Finally, we are aware of the limitation of this study concerning the size of the test sets---our current efforts are focused on validating the proposed framework over the scans acquired in other clinical sites (and using different scanners as well). We do, however, believe that the current validation procedure indeed provides solid insights that reflect the operational capabilities of the proposed technique.

\section*{Declaration of competing interest}
All authors declare that they have no competing interests.

\section*{Acknowledgments}
This work was supported by the National Centre for Research and Development POIR.01.01.01-00-0092/20. Jakub Nalepa was supported by the Silesian University of Technology grant for maintaining and developing research potential.

\section*{Supplementary material}
A subset of 6 manually-registered \textit{test-GT} samples generated based on the scans belonging to the OrCaScore dataset~\cite{wolterink_evaluation_2016} and the corresponding segmentations using the non-contrast coronary artery segmentation model presented in the paper is available under: https://doi.org/10.5281/zenodo.7808199.



\bibliographystyle{elsarticle-num} 
\bibliography{CoronariesNCmod}

\begin{thebibliography}{10}
\expandafter\ifx\csname url\endcsname\relax
  \def\url#1{\texttt{#1}}\fi
\expandafter\ifx\csname urlprefix\endcsname\relax\def\urlprefix{URL }\fi
\expandafter\ifx\csname href\endcsname\relax
  \def\href#1#2{#2} \def\path#1{#1}\fi

\bibitem{go_executive_2013}
A.~S. Go, et~al., Executive {Summary}: {Heart} {Disease} and {Stroke}
  {Statistics}—2013 {Update}, Circ. 127~(1) (2013) 143--152.
\newblock \href {https://doi.org/10.1161/CIR.0b013e318282ab8f}
  {\path{doi:10.1161/CIR.0b013e318282ab8f}}.

\bibitem{sun_computational_2014}
Z.~Sun, L.~Xu, Computational fluid dynamics in coronary artery disease, Comput.
  Med. Imaging Graph. 38~(8) (2014) 651--663.
\newblock \href {https://doi.org/10.1016/j.compmedimag.2014.09.002}
  {\path{doi:10.1016/j.compmedimag.2014.09.002}}.

\bibitem{malawski_deep_2022}
F.~Malawski, et~al., Deep {Learning} {Meets} {Computational} {Fluid} {Dynamics}
  to {Assess} {CAD} in {CCTA}, in: S.~Wu, B.~Shabestari, L.~Xing (Eds.),
  Applications of {Medical} {Artificial} {Intelligence}, Lecture {Notes} in
  {Computer} {Science}, Springer Nature Switzerland, Cham, 2022, pp. 8--17.
\newblock \href {https://doi.org/10.1007/978-3-031-17721-7\_2}
  {\path{doi:10.1007/978-3-031-17721-7\_2}}.

\bibitem{beier_hemodynamics_2016}
S.~Beier, et~al., Hemodynamics in {Idealized} {Stented} {Coronary} {Arteries}:
  {Important} {Stent} {Design} {Considerations}, Ann. Biomed. Eng. 44~(2)
  (2016) 315--329.
\newblock \href {https://doi.org/10.1007/s10439-015-1387-3}
  {\path{doi:10.1007/s10439-015-1387-3}}.

\bibitem{sun_personalized_2019}
Z.~Sun, S.~Jansen, Personalized {3D} printed coronary models in coronary
  stenting, Quant. Imaging. Med. Surg. 9~(8) (2019) 1356--1367.
\newblock \href {https://doi.org/10.21037/qims.2019.06.21}
  {\path{doi:10.21037/qims.2019.06.21}}.

\bibitem{silva_emerging_2018}
J.~N. Silva, M.~Southworth, C.~Raptis, J.~Silva, Emerging {Applications} of
  {Virtual} {Reality} in {Cardiovascular} {Medicine}, JACC: Basic Transl. Sci.
  3~(3) (2018) 420--430.
\newblock \href {https://doi.org/10.1016/j.jacbts.2017.11.009}
  {\path{doi:10.1016/j.jacbts.2017.11.009}}.

\bibitem{yoo_hands-surgical_2017}
S.-J. Yoo, T.~Spray, E.~H. Austin, T.-J. Yun, G.~S. van Arsdell, Hands-on
  surgical training of congenital heart surgery using 3-dimensional print
  models, J. Thorac. Cardiovasc. Surg. 153~(6) (2017) 1530--1540.
\newblock \href {https://doi.org/10.1016/j.jtcvs.2016.12.054}
  {\path{doi:10.1016/j.jtcvs.2016.12.054}}.

\bibitem{isensee_nnu-net_2021}
F.~Isensee, P.~F. Jaeger, S.~A.~A. Kohl, J.~Petersen, K.~H. Maier-Hein,
  {nnU}-{Net}: a self-configuring method for deep learning-based biomedical
  image segmentation, Nat. Methods 18~(2) (2021) 203--211.
\newblock \href {https://doi.org/10.1038/s41592-020-01008-z}
  {\path{doi:10.1038/s41592-020-01008-z}}.

\bibitem{wasserthal_totalsegmentator_2022}
J.~Wasserthal, M.~Meyer, H.-C. Breit, J.~Cyriac, S.~Yang, M.~Segeroth,
  {TotalSegmentator}: robust segmentation of 104 anatomical structures in {CT}
  images, arXiv:2208.05868 (2022).

\bibitem{gharleghi_automated_2022}
R.~Gharleghi, et~al., Automated segmentation of normal and diseased coronary
  arteries – {The} {ASOCA} challenge, Comput. Med. Imaging Graph. 97 (2022)
  102049.
\newblock \href {https://doi.org/10.1016/j.compmedimag.2022.102049}
  {\path{doi:10.1016/j.compmedimag.2022.102049}}.

\bibitem{siddique_u-net_2021}
N.~Siddique, S.~Paheding, C.~P. Elkin, V.~Devabhaktuni, U-{Net} and {Its}
  {Variants} for {Medical} {Image} {Segmentation}: {A} {Review} of {Theory} and
  {Applications}, IEEE Access 9 (2021) 82031--82057.
\newblock \href {https://doi.org/10.1109/ACCESS.2021.3086020}
  {\path{doi:10.1109/ACCESS.2021.3086020}}.

\bibitem{Shit_2021_CVPR}
S.~Shit, et~al., {clDice - A Novel Topology-Preserving Loss Function for
  Tubular Structure Segmentation}, in: Proceedings of the IEEE/CVF Conference
  on Computer Vision and Pattern Recognition (CVPR), 2021, pp.
  {16560}--{16569}.

\bibitem{agatston_quantification_1990}
A.~S. Agatston, W.~R. Janowitz, F.~J. Hildner, N.~R. Zusmer, M.~Viamonte,
  R.~Detrano, Quantification of coronary artery calcium using ultrafast
  computed tomography, J. Am. Coll. Cardiol. 15~(4) (1990) 827--832.
\newblock \href {https://doi.org/10.1016/0735-1097(90)90282-T}
  {\path{doi:10.1016/0735-1097(90)90282-T}}.

\bibitem{isgum_automatic_2012}
I.~Isgum, M.~Prokop, M.~Niemeijer, M.~A. Viergever, B.~van Ginneken, Automatic
  {Coronary} {Calcium} {Scoring} in {Low}-{Dose} {Chest} {Computed}
  {Tomography}, IEEE Trans. Med. Imaging 31~(12) (2012) 2322--2334.
\newblock \href {https://doi.org/10.1109/TMI.2012.2216889}
  {\path{doi:10.1109/TMI.2012.2216889}}.

\bibitem{wolterink_automatic_2014}
J.~M. Wolterink, T.~Leiner, R.~A.~P. Takx, M.~A. Viergever, I.~Išgum, An
  automatic machine learning system for coronary calcium scoring in clinical
  non-contrast enhanced, {ECG}-triggered cardiac {CT}, in: Medical Imaging
  2014: Computer-Aided Diagnosis, San Diego, California, USA, 2014, p. 90350E.
\newblock \href {https://doi.org/10.1117/12.2042226}
  {\path{doi:10.1117/12.2042226}}.

\bibitem{yang_automatic_2016}
G.~Yang, Y.~Chen, X.~Ning, Q.~Sun, H.~Shu, J.-L. Coatrieux, Automatic coronary
  calcium scoring using noncontrast and contrast {CT} images: {Automatic}
  coronary calcium scoring, Med. Phys. 43~(5) (2016) 2174--2186.
\newblock \href {https://doi.org/10.1118/1.4945045}
  {\path{doi:10.1118/1.4945045}}.

\bibitem{lessmann_automatic_2018}
N.~Lessmann, et~al., Automatic {Calcium} {Scoring} in {Low}-{Dose} {Chest} {CT}
  {Using} {Deep} {Neural} {Networks} {With} {Dilated} {Convolutions}, IEEE
  Trans. Med. Imaging 37~(2) (2018) 615--625.
\newblock \href {https://doi.org/10.1109/TMI.2017.2769839}
  {\path{doi:10.1109/TMI.2017.2769839}}.

\bibitem{shahzad_patient-specific_2010}
R.~Shahzad, et~al., A patient-specific coronary density estimate, in: 2010
  {IEEE} {International} {Symposium} on {Biomedical} {Imaging}: {From} {Nano}
  to {Macro}, IEEE, Rotterdam, Netherlands, 2010, pp. 9--12.
\newblock \href {https://doi.org/10.1109/ISBI.2010.5490426}
  {\path{doi:10.1109/ISBI.2010.5490426}}.

\bibitem{klein_elastix_2010}
S.~Klein, M.~Staring, K.~Murphy, M.~Viergever, J.~Pluim, elastix: {A} {Toolbox}
  for {Intensity}-{Based} {Medical} {Image} {Registration}, IEEE Trans. Med.
  Imaging 29~(1) (2010) 196--205.
\newblock \href {https://doi.org/10.1109/TMI.2009.2035616}
  {\path{doi:10.1109/TMI.2009.2035616}}.

\bibitem{shamonin_fast_2013}
D.~Shamonin, Fast parallel image registration on {CPU} and {GPU} for diagnostic
  classification of {Alzheimer}'s disease, Front. Neuroinform. 7 (2013).
\newblock \href {https://doi.org/10.3389/fninf.2013.00050}
  {\path{doi:10.3389/fninf.2013.00050}}.

\bibitem{joshi_unbiased_2004}
S.~Joshi, B.~Davis, M.~Jomier, G.~Gerig, Unbiased diffeomorphic atlas
  construction for computational anatomy, NeuroImage 23 (2004) S151--S160.
\newblock \href {https://doi.org/10.1016/j.neuroimage.2004.07.068}
  {\path{doi:10.1016/j.neuroimage.2004.07.068}}.

\bibitem{shahzad_vessel_2013}
R.~Shahzad, et~al., Vessel {Specific} {Coronary} {Artery} {Calcium} {Scoring},
  Acad. Radiol. 20~(1) (2013) 1--9.
\newblock \href {https://doi.org/10.1016/j.acra.2012.07.018}
  {\path{doi:10.1016/j.acra.2012.07.018}}.

\bibitem{langerak_label_2010}
T.~R. Langerak, U.~A. van~der Heide, A.~N. T.~J. Kotte, M.~A. Viergever, M.~van
  Vulpen, J.~P.~W. Pluim, Label {Fusion} in {Atlas}-{Based} {Segmentation}
  {Using} a {Selective} and {Iterative} {Method} for {Performance} {Level}
  {Estimation} ({SIMPLE}), IEEE Trans. Med. Imaging 29~(12) (2010) 2000--2008.
\newblock \href {https://doi.org/10.1109/TMI.2010.2057442}
  {\path{doi:10.1109/TMI.2010.2057442}}.

\bibitem{kelm_automatic_2014}
B.~M. Kelm, Y.~Zheng, Automatic {Coronary} {Calcium} {Scoring} {Using} {Native}
  and {Contrasted} {CT} {Acquisitions}, MICCAI Challenge on Automatic Coronary
  Calcium Scoring (2014).

\bibitem{zheng_robust_2013}
Y.~Zheng, H.~Tek, G.~Funka-Lea, Robust and {Accurate} {Coronary} {Artery}
  {Centerline} {Extraction} in {CTA} by {Combining} {Model}-{Driven} and
  {Data}-{Driven} {Approaches}, in: Advanced {Information} {Systems}
  {Engineering}, Vol. 7908, Springer, Berlin, Heidelberg, 2013, pp. 74--81.
\newblock \href {https://doi.org/10.1007/978-3-642-40760-4\_10}
  {\path{doi:10.1007/978-3-642-40760-4\_10}}.

\bibitem{zheng_four-chamber_2008}
Y.~Zheng, A.~Barbu, B.~Georgescu, M.~Scheuering, D.~Comaniciu, Four-{Chamber}
  {Heart} {Modeling} and {Automatic} {Segmentation} for 3-{D} {Cardiac} {CT}
  {Volumes} {Using} {Marginal} {Space} {Learning} and {Steerable} {Features},
  IEEE Trans. Med. Imaging 27~(11) (2008) 1668--1681.
\newblock \href {https://doi.org/10.1109/TMI.2008.2004421}
  {\path{doi:10.1109/TMI.2008.2004421}}.

\bibitem{zheng_pericardium_2014}
Y.~Zheng, Pericardium {Based} {Model} {Fusion} of {CT} and {Non}-contrasted
  {C}-arm {CT} for {Visual} {Guidance} in {Cardiac} {Interventions}, in:
  Medical {Image} {Computing} and {Computer}-{Assisted} {Intervention} –
  {MICCAI} 2014, Vol. 8674 of Lecture Notes in Computer Science, Springer
  International Publishing, Cham, 2014, pp. 700--707.
\newblock \href {https://doi.org/10.1007/978-3-319-10470-6\_87}
  {\path{doi:10.1007/978-3-319-10470-6\_87}}.

\bibitem{kondo_semi-automatic_2015}
S.~Kondo, Semi-{Automatic} {Detection} of {Coronary} {Artery} {Calcium} with an
  {Artery} {Identiﬁcation} {Technique} (2015).

\bibitem{johnson_brainsfit_2007}
H.~Johnson, G.~Harris, K.~Williams, {BRAINSFit}: {Mutual} {Information}
  {Registrations} of {Whole}-{Brain} {3D} {Images}, {Using} the {Insight}
  {Toolkit}, The Insight Journal 180 (2007).
\newblock \href {https://doi.org/10.54294/hmb052} {\path{doi:10.54294/hmb052}}.

\bibitem{frangi_multiscale_1998}
A.~F. Frangi, W.~J. Niessen, K.~L. Vincken, M.~A. Viergever, Multiscale vessel
  enhancement filtering, in: W.~M. Wells, A.~Colchester, S.~Delp (Eds.),
  Medical {Image} {Computing} and {Computer}-{Assisted} {Intervention} —
  {MICCAI}’98, Lecture {Notes} in {Computer} {Science}, Springer, Berlin,
  Heidelberg, 1998, pp. 130--137.
\newblock \href {https://doi.org/10.1007/BFb0056195}
  {\path{doi:10.1007/BFb0056195}}.

\bibitem{lee_fully_2021}
J.-G. Lee, et~al., Fully {Automatic} {Coronary} {Calcium} {Score} {Software}
  {Empowered} by {Artificial} {Intelligence} {Technology}: {Validation} {Study}
  {Using} {Three} {CT} {Cohorts}, Korean J. Radiol. 22~(11) (2021) 1764.
\newblock \href {https://doi.org/10.3348/kjr.2021.0148}
  {\path{doi:10.3348/kjr.2021.0148}}.

\bibitem{suganyadevi_review_2022}
S.~Suganyadevi, V.~Seethalakshmi, K.~Balasamy, A review on deep learning in
  medical image analysis, Int. J. Multimed. Inf. Retr. 11~(1) (2022) 19--38.
\newblock \href {https://doi.org/10.1007/s13735-021-00218-1}
  {\path{doi:10.1007/s13735-021-00218-1}}.

\bibitem{chen_recent_2022}
X.~Chen, et~al., Recent advances and clinical applications of deep learning in
  medical image analysis, Med. Image Anal. 79 (2022) 102444.
\newblock \href {https://doi.org/10.1016/j.media.2022.102444}
  {\path{doi:10.1016/j.media.2022.102444}}.

\bibitem{kotowski_detecting_2023}
K.~Kotowski, et~al., Detecting liver cirrhosis in computed tomography scans
  using clinically-inspired and radiomic features, Comput. Biol. Med. 152
  (2023) 106378.
\newblock \href {https://doi.org/10.1016/j.compbiomed.2022.106378}
  {\path{doi:10.1016/j.compbiomed.2022.106378}}.

\bibitem{brunner_toward_2010}
G.~Brunner, D.~R. Chittajallu, U.~Kurkure, I.~A. Kakadiaris, Toward the
  automatic detection of coronary artery calcification in non-contrast computed
  tomography data, Int. J. Card. Imaging 26~(7) (2010) 829--838.
\newblock \href {https://doi.org/10.1007/s10554-010-9608-1}
  {\path{doi:10.1007/s10554-010-9608-1}}.

\bibitem{gonzalez_automated_2016}
G.~Gonzalez, G.~R. Washko, R.~S.~J. Estepar, Automated {Agatston} score
  computation in a large dataset of non {ECG}-gated chest computed tomography,
  in: 2016 {IEEE} 13th {International} {Symposium} on {Biomedical} {Imaging}
  ({ISBI}), IEEE, Prague, Czech Republic, 2016, pp. 53--57.
\newblock \href {https://doi.org/10.1109/ISBI.2016.7493209}
  {\path{doi:10.1109/ISBI.2016.7493209}}.

\bibitem{simonyan_very_2015}
K.~Simonyan, A.~Zisserman, Very {Deep} {Convolutional} {Networks} for
  {Large}-{Scale} {Image} {Recognition}, arXiv:1409.1556 (2015).
\newblock \href {https://doi.org/10.48550/arXiv.1409.1556}
  {\path{doi:10.48550/arXiv.1409.1556}}.

\bibitem{kurkure_supervised_2010}
U.~Kurkure, D.~R. Chittajallu, G.~Brunner, Y.~H. Le, I.~A. Kakadiaris, A
  supervised classification-based method for coronary calcium detection in
  non-contrast {CT}, Int. J. Card. Imaging 26~(7) (2010) 817--828.
\newblock \href {https://doi.org/10.1007/s10554-010-9607-2}
  {\path{doi:10.1007/s10554-010-9607-2}}.

\bibitem{wolterink_automatic_2015}
J.~M. Wolterink, T.~Leiner, R.~A.~P. Takx, M.~A. Viergever, I.~Isgum, Automatic
  {Coronary} {Calcium} {Scoring} in {Non}-{Contrast}-{Enhanced}
  {ECG}-{Triggered} {Cardiac} {CT} {With} {Ambiguity} {Detection}, IEEE Trans.
  Med. Imaging 34~(9) (2015) 1867--1878.
\newblock \href {https://doi.org/10.1109/TMI.2015.2412651}
  {\path{doi:10.1109/TMI.2015.2412651}}.

\bibitem{Jiang2022}
X.-Y. Jiang, Z.-Q. Shao, Y.-T. Chai, Y.-N. Liu, Y.~Li, Non-contrast {CT}-based
  radiomic signature of pericoronary adipose tissue for screening non-calcified
  plaque, Phys. Med. Biol. 67~(10) (2022) 105004.
\newblock \href {https://doi.org/10.1088/1361-6560/ac69a7}
  {\path{doi:10.1088/1361-6560/ac69a7}}.

\bibitem{yushkevich_fast_2016}
P.~A. Yushkevich, J.~Pluta, H.~Wang, L.~E. Wisse, S.~Das, D.~Wolk, Fast
  {Automatic} {Segmentation} of {Hippocampal} {Subfields} and {Medial}
  {Temporal} {Lobe} {Subregions} {In} 3 {Tesla} and 7 {Tesla} {T2}-{Weighted}
  {MRI}, Alzheimer's \& Dementia 12~(7) (2016) P126--P127.
\newblock \href {https://doi.org/10.1016/j.jalz.2016.06.205}
  {\path{doi:10.1016/j.jalz.2016.06.205}}.

\bibitem{keszei_survey_2017}
A.~P. Keszei, B.~Berkels, T.~M. Deserno, Survey of {Non}-{Rigid} {Registration}
  {Tools} in {Medicine}, J. Digit. Imaging 30~(1) (2017) 102--116.
\newblock \href {https://doi.org/10.1007/s10278-016-9915-8}
  {\path{doi:10.1007/s10278-016-9915-8}}.

\bibitem{lambert_segthor_2020}
Z.~Lambert, C.~Petitjean, B.~Dubray, S.~Kuan, {SegTHOR}: {Segmentation} of
  {Thoracic} {Organs} at {Risk} in {CT} images, in: 2020 {Tenth}
  {International} {Conference} on {Image} {Processing} {Theory}, {Tools} and
  {Applications} ({IPTA}), 2020, pp. 1--6, iSSN: 2154-512X.
\newblock \href {https://doi.org/10.1109/IPTA50016.2020.9286453}
  {\path{doi:10.1109/IPTA50016.2020.9286453}}.

\bibitem{saha_survey_2016}
P.~K. Saha, G.~Borgefors, G.~Sanniti~di Baja, A survey on skeletonization
  algorithms and their applications, Pattern Recognit. Lett. 76 (2016) 3--12.
\newblock \href {https://doi.org/10.1016/j.patrec.2015.04.006}
  {\path{doi:10.1016/j.patrec.2015.04.006}}.

\bibitem{izzo_vascular_2018}
R.~Izzo, D.~Steinman, S.~Manini, L.~Antiga, The {Vascular} {Modeling}
  {Toolkit}: {A} {Python} {Library} for the {Analysis} of {Tubular}
  {Structures} in {Medical} {Images}, J. Open Source Softw. 3~(25) (2018) 745.
\newblock \href {https://doi.org/10.21105/joss.00745}
  {\path{doi:10.21105/joss.00745}}.

\bibitem{wolterink_evaluation_2016}
J.~M. Wolterink, et~al., An evaluation of automatic coronary artery calcium
  scoring methods with cardiac {CT} using the {orCaScore} framework:
  {Evaluation} of cardiac {CT}-based automatic coronary calcium scoring, Med.
  Phys. 43~(5) (2016) 2361--2373.
\newblock \href {https://doi.org/10.1118/1.4945696}
  {\path{doi:10.1118/1.4945696}}.

\bibitem{bujny_dataset_2023}
M.~Bujny, et~al., \href{https://doi.org/10.5281/zenodo.7808199}{{{Data from:
  Coronary artery segmentation in non-contrast calcium scoring CT images using
  deep learning}}} (2023).
\newblock \href {https://doi.org/10.5281/zenodo.7808199}
  {\path{doi:10.5281/zenodo.7808199}}.
\newline\urlprefix\url{https://doi.org/10.5281/zenodo.7808199}

\end{thebibliography}





\end{document}